%% file: text-working.tex
\documentclass[preprint,review,12pt,authoryear]{elsarticle}

\usepackage[top=2cm, bottom=2cm, left=2.5cm, right=2.5cm]{geometry}

\usepackage{ragged2e}
\justifying
\usepackage{url}
\usepackage{amssymb}
\usepackage{amsmath}
\usepackage{gensymb}

\usepackage[inline]{trackchanges} 
\usepackage{booktabs} 
\usepackage{threeparttable} 
\usepackage{lmodern}

\journal{Marine Pollution Bulletin}

\begin{document}

\begin{frontmatter}



\title{Effect of Biofouling on Microplastic Transport in a 3-D Global Eulerian Model} 


\author[label1]{Zih-En Tseng\corref{cor1}} 
\author[label1]{Yue Wu}
\author[label2]{Chris Ruf}
\author[label3]{Dimitris Menemenlis}
\author[label1]{Yulin Pan}

\cortext[cor1]{Corresponding author}

\affiliation[label1]{organization={Department of Naval Architecture \& Marine engineering, the University of Michigan--Ann Arbor},
            addressline={1109 Geddes Ave}, 
            city={Ann Arbor},
            postcode={48109}, 
            state={Michigan},
            country={United States}}
            
\affiliation[label2]{organization={Department of Climate and Space Sciences and Engineering, the University of Michigan--Ann Arbor},
            addressline={1109 Geddes Ave}, 
            city={Ann Arbor},
            postcode={48109}, 
            state={Michigan},
            country={United States}}

\affiliation[label3]{organization={Moss Landing Marine Laboratories, San Jose State University},
            addressline={8272 Moss Landing Road}, 
            city={Moss Landing},
            postcode={95039}, 
            state={California},
            country={United States}}

\begin{abstract}
Biofouling---the occupation of microplastic (MP) surfaces by marine microbes---alters particles' buoyancy and transport, yet its effect on the global distribution of MPs has not been well quantified. We present the first three-dimensional global Eulerian model to fully couple MP transport with biofouling, by augmenting the concentration field with an extra dimension representing the biomass attachment density on MP surfaces. This approach embeds time-dependent particle properties directly into the Eulerian concentration field, overcoming a fundamental challenge of tracking property evolution in grid-based models. Idealized simulations show that biofouling significantly reshapes the vertical distribution of MPs when two conditions are met: the particles must be sufficiently buoyant when they are clean to remain near the sea surface, and the local plankton growth rate must exceed the decay rate. In three-dimensional global simulations, biofouling substantially alters the distribution of large MPs ($\gtrsim 10$ $\mu$m): biofouled particles are transported below the mixed layer to 500 m depth, and the subtropical surface garbage patches become more dispersed with reduced peak concentrations. This dispersion is due to a subsurface transport route, where biofouled particles sink into layers with reversed current and are carried outward from the gyre centers before regaining buoyancy. Small particles ($\lesssim 1$ $\mu$m) remain unaffected as they stay effectively neutrally buoyant even when biofouled. A comparison with a global trawler dataset shows that incorporating biofouling reduces the fraction of outlying model-observation data points from 25\% to 13\%, demonstrating a meaningful improvement in model skill. \\ \\
\end{abstract}


\begin{highlights}
\item First Eulerian model to couple global microplastic transport with biofouling
\item Biofouling disperses subtropical garbage patches via a subsurface transport route
\item Large particles ($\gtrsim 10$ $\mu$m) are transported to depths of 500 m by biofouling
\item Biofouling halves model-observation outliers, improving agreement with trawler data
\end{highlights}


\end{frontmatter}



\section{Introduction}
\label{Intro}

Every year, up to 27.6 million tons of plastic waste enter the ocean through rivers, coastlines, and industrial vessel routes \citep{watt_ocean_2021,lau_evaluating_2020}. Once in the ocean, plastics are broken down by physical abrasion and chemical weathering into smaller and smaller fragments \citep{pfohl_environmental_2022}, eventually reaching the scale of microplastics (MPs)---particles smaller than 5 mm \citep{program_microplastics_2024}. This fragmentation process produces a broad size distribution of MPs, spanning from several nanometers to millimeters, and the resulting particles are found at all depths of the ocean \citep{ten_hietbrink_nanoplastic_2025,pabortsava_high_2020}. Unlike large debris such as fishing nets that can be visually identified, MPs are small enough to enter the food chain invisibly, accumulate in animal tissue, and even pass through the blood-brain barrier \citep{nihart_bioaccumulation_2025}. Furthermore, MPs that enter the atmosphere through wind-driven re-suspension and fragmentation contribute to atmospheric warming by absorbing solar radiation \citep{liu_atmospheric_2026}. These properties make MPs a threat to marine life and human health, with impacts on fishery sustainability, water quality, and climate change \citep{vazquez-rowe_microplastics_2021}. Knowing where MPs accumulate in the ocean is essential for assessing these risks and designing effective pollution management strategies.

Field measurements offer direct observation of the distribution of oceanic MPs. Trawler nets collect particles larger than $\sim 300\ \mu$m, and finer filtering systems capture particles down to $\sim 10\ \mu$m \citep{lindeque_are_2020,nyadjro_noaa_2023,enders_abundance_2015}. Together, these measurements have established that MPs accumulate at the sea surface in the subtropical gyres \citep{eriksen_plastic_2014,cozar_plastic_2014}. More recently, water samples collected at multiple depths have revealed a vertical distribution of MPs \citep{zhao_distribution_2025,ten_hietbrink_nanoplastic_2025,pabortsava_high_2020}. Still, these field measurements suffer from limitations: particles smaller than the detection threshold are systematically underrepresented, and the collected data are sparse in space and time. Remote sensing provides indirect observation of the distribution of MPs via sea surface roughness along satellite ground tracks \citep{jones-williams_remote_2021,Maddy_Chris_2022}, which covers almost the global oceans except the polar regions. However, results must be interpreted with caution because they rely on retrieval algorithms, and the derived MP concentrations remain sensitive to atmospheric and sea surface conditions such as heavy precipitation.

Numerical modeling serves as a valuable complement to the above observations, with simulation approaches generally categorized into Lagrangian and Eulerian modeling. Lagrangian models track the location of individual particles to resolve their trajectories. Eulerian models, on the other hand, solve for the evolution of particle concentration on a fixed grid. Compared to Lagrangian models, the Eulerian approach has three advantages \citep{tseng_distribution_2025}: (1) the computational cost does not depend on the number of particles; (2) mature parametrization schemes are available to represent subgrid-scale processes such as vertical mixing; and (3) data assimilation can be accommodated in a straight-forward manner.

Here we summarize previous findings on MP distribution from global Lagrangian and Eulerian models. Earlier two-dimensional (2-D) Lagrangian models simplify the problem by treating MPs as surface tracers, and predict that buoyant particles accumulate in subtropical gyres \citep{lebreton_numerical_2012,chenillat_fate_2021}. Three-dimensional (3-D) models extend this picture by considering the vertical motion of particles, allowing the distinction between particles with different properties \citep[e.g., density; ][]{lobelle_global_2021,mountford_eulerian_2019,richon_zooplankton_2022,richon_faecal_2026}. \citet{tseng_distribution_2025} further incorporates a more physical treatment of buoyancy that enables a distinction among MPs of different sizes. Together, these models have established that positively buoyant particles (i.e., with density lower than seawater and size above the threshold of 10 $\mu$m) aggregate within the mixed layer (ML) in subtropical gyres, while particles that are small enough (with size $\lesssim 1$ $\mu$m) can be treated as neutrally buoyant regardless of their material density and are transported deeper into the water column so do not form surface garbage patches in the gyres.


While most models discussed above assume that particle properties remain constant throughout the transport process, various oceanic mechanisms, such as biofouling, can significantly alter particle properties over time and subsequently affect their distribution. In the surface ocean, where microbial concentrations reach the order of $10^6$ cells ml$^{-1}$ \citep{zakem_redox-informed_2020}, microbes readily colonize MP surfaces and form a biofilm known as the plastisphere \citep{cable_distribution_2017}. As the biofilm grows, both the size and density of the particle increase, causing the particle to sink. Once a biofouled particle descends into deeper waters, the biofilm decays due to reduced sunlight availability and the particle regains buoyancy, resulting in oscillatory vertical motions \citep{olivia_microplastics_2026}. \citet{kooi_ups_2017} first formalized this process into a biofouling model, which was later extended to a global-scale simulation by \citet{lobelle_global_2021}. However, \citet{lobelle_global_2021} focused only on the timescale of vertical transport of particles without providing a global concentration map. As a result, the effect of biofouling on the global distribution of MPs remains unexplored.

Because of the aforementioned advantages of Eulerian models, it is worthwhile to build an Eulerian model that incorporates biofouling. However, this poses a fundamental challenge. In a Lagrangian model, the properties of individual particle can be tracked and updated as the particle moves in space and time. In an Eulerian model, in contrast, only the concentration field is simulated, with no explicit means to track the properties of individual particles. We address this difficulty by augmenting the model with an additional dimension ($\S$\ref{sec:transport-equations}) representing the amount of biomass attached on MP surfaces, effectively embedding the biofouling process into the concentration field. This yields the first Eulerian model to our knowledge to fully couple MP transport with biofouling on a global scale.

With our newly developed model, we first perform idealized 1-D simulations to study the vertical profile of biofouled particles. We demonstrate that biofouling can reshape the vertical distribution of MPs under suitable environmental conditions, corroborating a stability analysis we develop. We then perform 3-D global simulations to unravel the effect of biofouling on the global distribution pattern and transport of MPs. For large particles ($\gtrsim 10\ \mu$m), biofouling results in transport below the mixed layer down to $500$ m depth, in contrast to the non-biofouled case where particles aggregate within the mixed layer. The surface garbage patches in the subtropical gyres become more dispersed, showing a smoother surface concentration compared to the non-biofouled case. This dispersion is attributed to a subsurface transport, which moves biofouled MPs away from the centers of subtropical gyres. It reduces surface concentration variability, bringing simulation results into closer agreement with trawler measurements \citep{isobe_multilevel_2021} in most oceans. For sufficiently small particles ($\lesssim 1\ \mu$m), their concentration and transport remain insensitive to biofouling, as they stay neutrally buoyant despite being biofouled.

The paper is organized as follows: in $\S$\ref{sec:methodology} we explain our methodology and present model equations; in $\S$\ref{sec:exp_design} we present the experiment designs, discuss the results for both 1-D and 3-D simulations, and provide physical explanations for the findings; in $\S$\ref{sec:compare_obs} we validate the simulation results against a modern microplastic dataset from trawler measurements \citep{isobe_multilevel_2021} to demonstrate the improved agreement achieved when biofouling is incorporated.

\section{Methodology}
\label{sec:methodology}

We present the effect of biofouling on particle properties, followed by Kooi's equation governing the behavior of biofouled particles ($\S$\ref{sec:biofouling}). Next, we show the complete transport equation of MPs ($\S$\ref{sec:transport-equations}), with an emphasis on the additional treatment to represent the biofouling process.

\subsection{Dynamic particle properties due to biofouling}
\label{sec:biofouling}
\begin{figure}[h!]
\centering
\includegraphics[width=.5\textwidth]{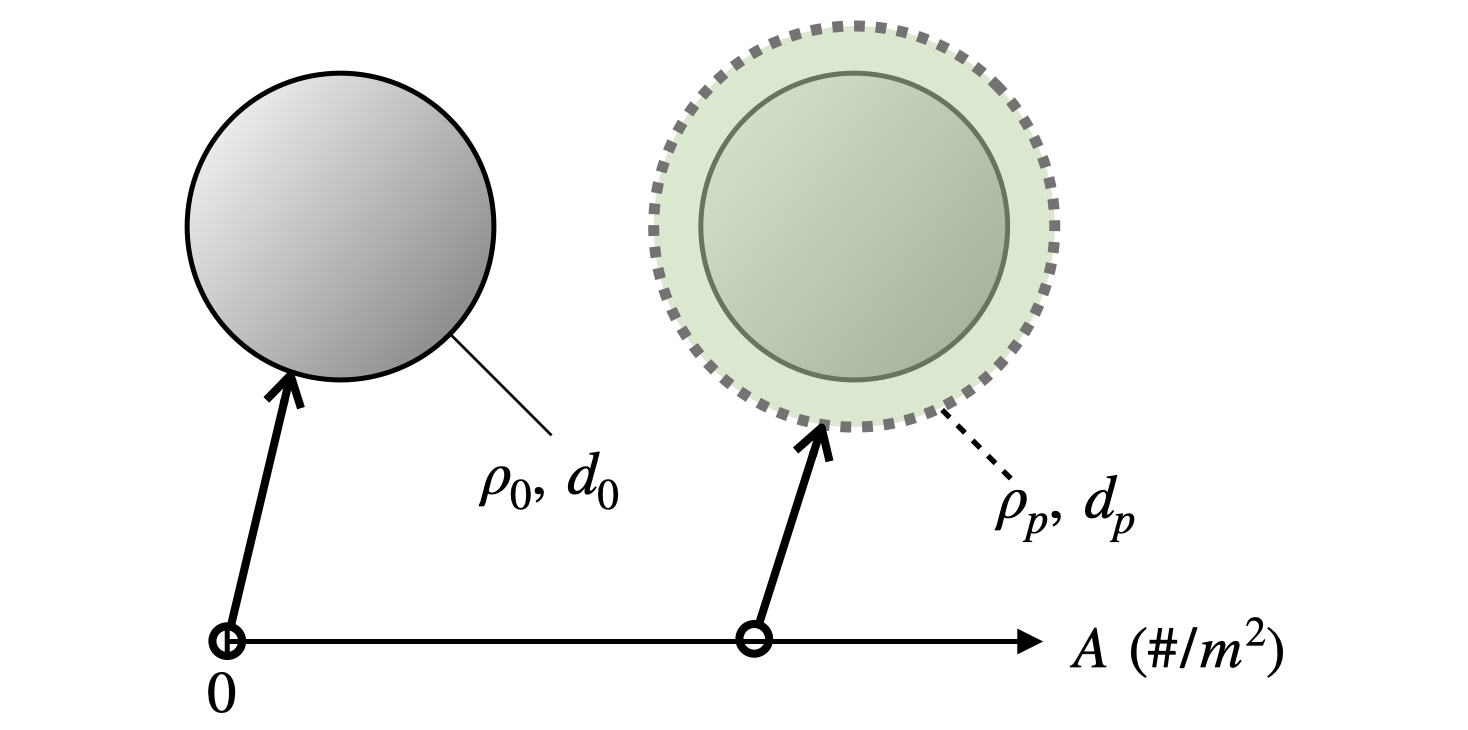}
\caption{Schematic of (left) the plastic core and (right) a biofouled particle.}\label{fig:attach_mp}
\end{figure}

Biofouling results in compound particles composed of both organic and inorganic materials. This process can be quantified by the time-dependent attachment density $A(t)$ in $\#$ m$^{-2}$, which represents the number of plankton cells attached per unit area of a particle's surface. It modifies particle's effective density $\rho_p$ and diameter $d_p$, which subsequently reshape the distribution of MPs by altering particle terminal velocity \citep{tseng_distribution_2025,dey_terminal_2019,khatmullina_settling_2017}
\begin{equation}
    w_r = \dfrac{g d_p^2 (\rho_w-\rho_p)}{18 \mu}, \label{eq:terminal_velocity}
\end{equation}
where $g$ is gravitational acceleration, $\rho_w$ is the water density, and $\mu$ is the dynamic viscosity. A schematic is provided in Figure \ref{fig:attach_mp} to differentiate a clean and a biofouled MP particle.

As the attachment $A(t)$ changes with time, the properties of a MP particle also vary
\begin{align}
    \begin{cases}
        d_p(t) = d_p(d_0, \rho_0, A(t)), \\
        \rho_p(t) = \rho_p(d_0, \rho_0, A(t)), \label{eq:attch_properties_0}
    \end{cases}
\end{align}
where $d_0$ and $\rho_0$ denote the diameter and density of a plastic core; $d_p$ and $\rho_p$ denote the effective diameter and density of a biofouled particle. Assuming that the plastic core of a particle is spherical and that the biofilm is evenly distributed over the particle's surface, $d_p$ and $\rho_p$ are determined by considering its compound volume and mass
\begin{equation}
    \begin{cases}
        \dfrac{4\pi}{3} \left( \dfrac{d_p}{2} \right) ^3 = \dfrac{4\pi}{3} \left( \dfrac{d_0}{2} \right) ^3 + 4\pi A \left( \dfrac{d_0}{2} \right) ^2 V_a, \\
        \dfrac{4\pi}{3} \left( \dfrac{d_p}{2} \right)^3 \rho_p = \dfrac{4\pi}{3} \left( \dfrac{d_0}{2} \right)^3 \rho_0 + 4\pi A \left( \dfrac{d_0}{2} \right)^2 V_a \rho_a, \label{eq:attch_properties_1}
    \end{cases}
\end{equation}
where the median density $\rho_a$ and volume $V_a$ of plankton cells are taken to be 1388 kg m$^{-3}$ and $2 \cdot 10^{-16}$ m$^3$, respectively \citep{fisher_interactions_1983,kooi_ups_2017}. Solving \eqref{eq:attch_properties_1} yields
\begin{equation}
    \begin{cases}
        d_p = \sqrt[3]{d_0^3 + 6 A d_0^2 V_a}, \\
        \rho_p = \dfrac{ d_0 \rho_0 + A V_a \rho_a}{d_0 + 6 A V_a}. \label{eq:attch_properties_2}
    \end{cases}
\end{equation}

The time rate of change for $A(t)$ is described by the biofouling equation \citep{kooi_ups_2017}
\begin{equation}
A' \equiv \dfrac{dA}{dt} = \dfrac{\beta_a A_a}{\theta_p} + \mu_a(T,I) A - Q_{10}^{(T-20)/10} R_{20} A - m_a A - g_aA.
\label{eq:bio-kooi}
\end{equation}

On the right hand side of \eqref{eq:bio-kooi}, the first term represents a gain in the attachment, assuming that a particle coagulates with an algae cell when they collide in Brownian motions. Here, $\beta_a$ is the encounter kernel rate in m$^3$ s$^{-1}$, $A_a$ is the ambient algae cell concentration in \# m$^{-3}$; and $\theta_p$ is the surface area in m$^2$ of the spherical MP particle. The second term represents the growth of algae, which relies on photosynthesis \citep{bernard_validation_2012}. Here, $\mu_a$ is the algae growth rate in $s^{-1}$, which depends on $T$, the seawater temperature, in $\degree$C and $I$, the shortwave solar irradiation, in $\mu$E m$^{-2}$ day$^{-1}$. The third term represents the decrease of algae due to metabolism, where $R_{20}$ is the plankton metabolism rate at 20$\degree$C in s$^{-1}$, and $Q_{10}$ is a unit-less constant modeling the dependency of plankton metabolism rate on temperature. The fourth term represents the decrease of algae due to their natural mortality, with $m_a$ being the mortality rate in s$^{-1}$. The last term represents the decrease of algae due to zooplankton grazing, with $g_a$ being the grazing rate in s$^{-1}$. The final three terms that have negative effects on $A'$ are encapsulated in the total decay rate $\gamma = Q_{10}^{(T-20)/10} R_{20}+m_a+g_a$. The parameters $Q_{10}$, $R_{20}$, $m_a$, and $g_a$ are collectively referred to as the ``biogeochemical parameters'' throughout the paper, and the choice of their values (as well as other parameters) will be discussed in $\S$\ref{sec:exp_design} along with the settings of our simulations.

Before transitioning to the full model transport equation, we present some mathematical properties of \eqref{eq:bio-kooi} and their associated physical significance. We note that \eqref{eq:bio-kooi} can be formulated as an ordinary differential equation with periodic coefficients, exhibiting diurnal variations in $\mu_a$, primarily in response to shortwave irradiation $I$. Following a standard Floquet stability analysis in \ref{append:floquet}, we find that the stability of \eqref{eq:bio-kooi} is characterized by the Floquet exponent
\begin{equation}
    q = \int_t^{t+T_d} \big[ \mu_a(I(t'),T) -m_a-g_a-Q_{10}^{(T-20)/10} R_{20} \big] \,dt',
    \label{eq:floquet-preview}
\end{equation}
where $T_d=1$ day. If a particle was held stationary, $q>0$ means that the solution $A(t)$ is unstable, and the attached algae can grow indefinitely. When $q<0$, it means the solution $A(t)$ is Lyapunov stable, and the amount of attachment remains finite. Therefore, $q>0$ within some depth is a necessary condition for biofouling to make a significant difference in the profile of MPs.

\subsection{Incorporation of biofouling into the transport equation}
\label{sec:transport-equations}

To incorporate biofouling into our Eulerian transport model, we must distinguish particles with different plastic cores and different attachment densities, since the three parameters ($\rho_0$, $d_0$, and $A$) together determine the effective density $\rho_p$ and diameter $d_p$. Therefore, for given $(\rho_0,d_0)$, we augment the grid space by spanning an additional dimension of $A$ to represent the extent of biofouling, and formulate the concentration of MPs on the 5-D space of $(x,y,z,t,A)$. To represent the biofouling process, we include an additional term, $Q_\text{biofouling}$ in model equation \citep{tseng_distribution_2025,zenodo_record_tseng2026}
\begin{equation}
\dfrac{\partial \tau}{\partial t} + \nabla \cdot (\tau \mathbf{u}) +\partial_z (\tau w_r) =\nabla \cdot (K \nabla \tau) + Q\, \delta (A) + Q_\text{biofouling},
\label{eq:adv-diff}
\end{equation}
\begin{equation}
Q_\text{biofouling} = - \dfrac{\partial (\tau A')}{\partial A},
\label{eq:bio_advect}
\end{equation}
where $\mathbf{u}$ and $K$ are velocity vector and diffusivity tensor from the ECCOv4r5 dataset \citep{ecco_consortium_synopsis_2021,ecco_consortium_ecco_2023,tseng_distribution_2025}, respectively, $Q$ is the coastal MP release rate based on \citet{jambeck_plastic_2015}, and $\delta(A)$ indicates that only clean particles are released. Except for the additional term and dimension, the equation \eqref{eq:adv-diff} is consistent with \citet{tseng_distribution_2025}, where readers can find detailed information of each term. In contrast to our previous model, the concentration $\tau$ is now $\tau(x,y,z,A,t)$ in unit $\#$ m$^{-3}$ ($\#$ m$^{-2}$)$^{-1}$.

\begin{figure}[h!]
\centering
\includegraphics[width=0.8\textwidth]{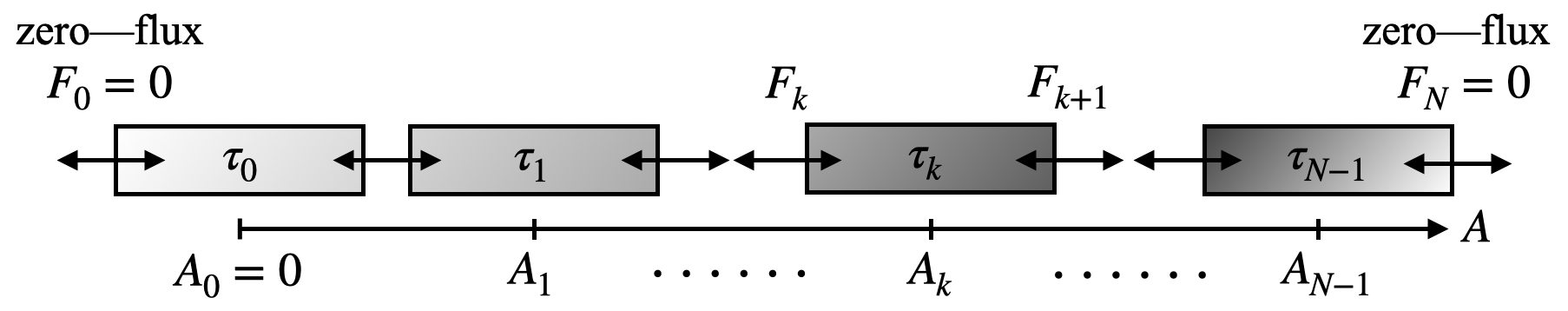}
\caption{Schematic of the distribution of particle concentration $\tau$ and flux $F$ along the dimension of $A$.}\label{fig:finite-volume}
\end{figure}

The biofouling process is formulated as an advective term along the dimension of $A$ \eqref{eq:bio_advect}, with $A'$ computed base on Kooi's biofouling equation \eqref{eq:bio-kooi}. The parameter $A$ is discretized into $N=40$ linearly spaced bins ranging from $[ 0, A_\text{max} ]$. Here, $A_\text{max} = 7.7 \times 10^{10}\#$ m$^{-2}$ is chosen based on empirical estimate of our numerical simulations, ensuring that particles do not artificially accumulate within the most biofouled bin. The divergence form of \eqref{eq:bio_advect} and the zero-flux boundary conditions together ensure the conservation of total number of particles $\iiiint \, \tau \,dx\,dy\,dz\,dA$ within the domain. A schematic is provided in Figure \ref{fig:finite-volume} to illustrate the discretization on the dimension of $A$, with the flux $F_k$ between cell $k-1$ and cell $k$ computed using the first-order upwind advection scheme
\begin{equation}
    F_k = 
    \begin{cases}
        0, & \quad \mbox{if} \quad k = 0 \mbox{ or } N,\\
        \tau(x,y,z,A_k)\, A'(x,y,z,\dfrac{A_{k-1}+A_k}{2}), & \quad \mbox{if} \quad A' \leq 0, \\
        \tau(x,y,z,A_{k-1})\, A'(x,y,z,\dfrac{A_{k-1}+A_k}{2}), & \quad \mbox{if} \quad A' > 0.
    \end{cases}
\end{equation}

\section{Experiment design and results}
\label{sec:exp_design}


\subsection{1-D cases}
\label{sec:exp_1D}

We first perform idealized simulations with only one spatial dimension of depth. The particles considered in the 1-D cases have cores with density $\rho_0=900$ kg m$^{-3}$ and size $d_0=50$ $\mu$m. Each simulation is run for 50 days under typical conditions of the subtropical North Pacific ocean in winter, with surface temperature being $25\degree$C and MLD being 80 m. The three simulations are labeled with their distinct settings related to biofouling as ``1D-CL'', ``1D-BF-L\'evy'', and ``1D-BF-Calbet'', respectively. In the case 1D-CL, particles are not biofouled and are always clean. In the cases 1D-BF-L\'evy and 1D-BF-Calbet, biofouling is considered and it introduces one additional dimension of attachment $A$. The two cases for biofouled particles differ by their choices of biogeochemical parameters, corresponding to \citet{levy_large-scale_2012} and \citet{calbet_phytoplankton_2004}. The settings of 1D-BF-Calbet differ from those of 1D-BF-L\'evy in that it has higher decay rate $\gamma$ through metabolism, mortality, and grazing \eqref{eq:bio-kooi}. Particularly, the case 1D-BF-Calbet is consistent with the experiment settings of \citet{kooi_ups_2017}, where the background algae $A_a(z)$, water temperature $T(z)$, salinity $S(z)$, water density $\rho_w(z)$, and viscosity $\mu(z)$ are taken from. The specific functional forms of the profiles and values of the parameters are documented in \ref{append:parameters}.

\begin{figure}[h!]
\centering
\includegraphics[width=1.\textwidth]{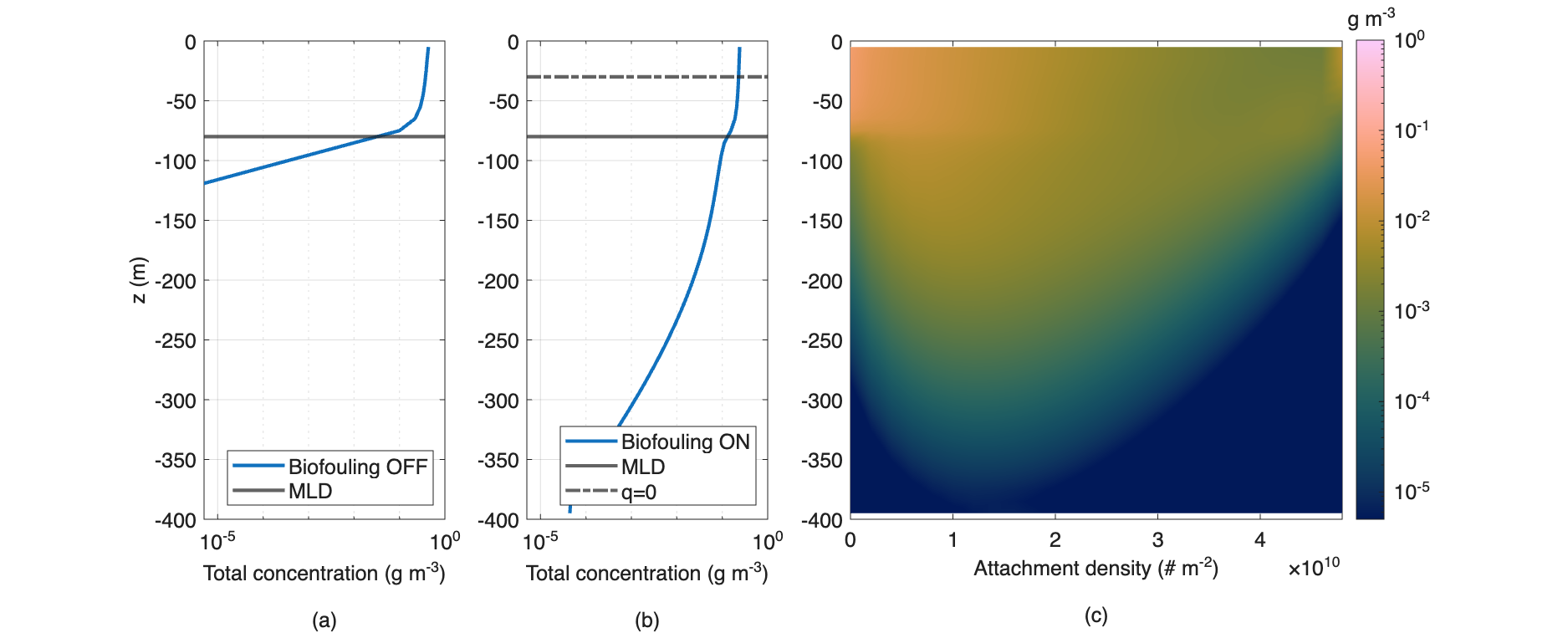}
\caption{The simulation results of (a) 1D-CL, showing the vertical profiles of non-biofouled particles; (b) 1D-BF-L\'evy, showing the vertical profiles of biofouled particles; and (c) 1D-BF-L\'evy, which is plotted in a joint space spanned by the depth and attachment density.}\label{fig:1D-levy}
\end{figure}

We now present the results from tests 1D-CL and 1D-BF-L\'evy, with their vertical profiles plotted in figure \ref{fig:1D-levy}a and figure \ref{fig:1D-levy}b, respectively. In the case 1D-CL without biofouling (figure \ref{fig:1D-levy}a), MPs show a homogeneous distribution within the ML, and the concentration diminishes rapidly below. The profiles of clean MPs we show here agrees with those reported in \citet{tseng_distribution_2025} and \citet{richon_zooplankton_2022}. In the case 1D-BF-L\'evy with biofouling (figure \ref{fig:1D-levy}b), the signatures of biofouled MPs penetrate deeper down to $\sim 400$ m as they become negatively buoyant. The deepening alters the vertical profile such that the concentration no longer diminishes at the bottom of the ML, and the surface concentration of MPs is reduced by $3$ times. In figure \ref{fig:1D-levy}c, we further show distribution of particles in the joint space of depth and attachment $A$, where we see larger values of $A$ for particles closer to the surface. This is because the low decay rate $\gamma$ from \citet{levy_large-scale_2012} allows the attachment to quickly grow, as indicated by the large values of Floquet exponent \eqref{eq:floquet-preview}. Moreover, as a particle gets biofouled, its high density forces it to sink down below the mixed layer. At $\gtrsim 50$ m depth where sunlight cannot reach, the growth rate approaches zero, the attachment starts to wane, and the Floquet exponent is dominated by $\gamma$. 

\begin{figure}[h!]
\centering
\includegraphics[width=1.\textwidth]{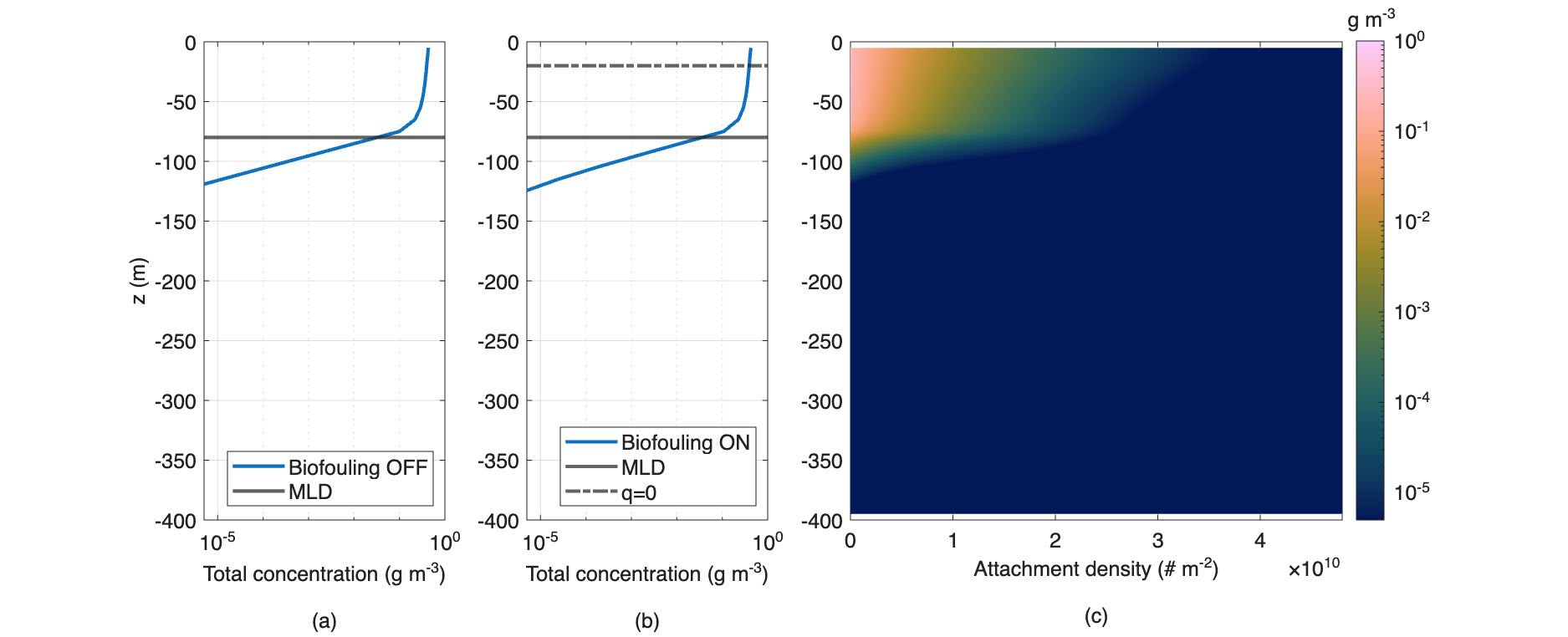}
\caption{The simulation results of (a) 1D-CL, showing the vertical profiles of non-biofouled particles; (b) 1D-BF-Calbet, showing the vertical profiles of biofouled particles; and (c) 1D-BF-Calbet, which is plotted in a joint space spanned by the depth and attachment density.}\label{fig:1D-calbet}
\end{figure}

We next show the results from test 1D-BF-Calbet. In figure \ref{fig:1D-calbet}, along with 1D-CL as a reference (figure \ref{fig:1D-calbet}a), we show our results of the case 1D-BF-Calbet. In terms of the vertical profile of biofouled particles (figure \ref{fig:1D-calbet}b), when higher decay rate $\gamma$ is used, the effect of biofouling is not significant. As shown in figure \ref{fig:1D-calbet}c, most particles aggregate within the mixed layer with low attachment density. This is because the higher $\gamma$ forces the biofilm to decay quickly, and the particle regains buoyancy before any significant vertical displacement occurs. The higher $\gamma$ is also reflected by the more negative Floquet exponent values throughout the water column. As a result, the vertical profile of biofouled MPs remains similar to its non-biofouled counterpart.

Although the setting of our simulation is consistent with \citet{kooi_ups_2017}, our result differs from their Lagrangian simulation, where the sign change of the Floquet exponent at $\sim 20$ m produces oscillatory motion of particles centered around that depth. Above this depth, net algae growth increases the particle density, while below it, net decay reduces the attachment and the particle becomes lighter. In our results, no such oscillation is observed. We attribute this difference to the vertical mixing within the mixed layer: the KPP-form diffusion redistributes particles across the mixed layer, smoothing out the oscillatory signal that would otherwise emerge. The Floquet exponent framework remains consistent with both results---the difference lies not in the biofouling dynamics, but in the treatment of vertical mixing.

\subsection{3-D cases}
\label{sec:exp_3D}

The following cases are ran on Texas A$\&$M High Performance Research Computing cluster with 96 CPUs and roughly 2 days of wall-time for each 25-year simulation. The simulations are performed on particles with low-density cores ($\rho_0=900$ kg m$^{-3}$) of three sizes: $d_0 = 50$, $10$, and $1$ $\mu$m. Each simulation is labeled as 3D-CL if particles are clean or 3D-BF if particles are biofouled, and the last two digits of numbers denote the diameter of the particle's core $d_0$. The discussion will be focused on the results of $50$ $\mu$m ($\S$\ref{sec:3D-50}) and $10$ $\mu$m ($\S$\ref{sec:3D-10}) particles. The results for the $1$ $\mu$m particles are briefly discussed in $\S$\ref{sec:3D-1}, as biofouling does not alter their distribution, because $1$ $\mu$m particles remain effectively neutrally buoyant \citep{tseng_distribution_2025}.

The global simulations are performed on the so-called llc-90 grid, which has the same spatial resolution as the ECCOv4r5 dataset \citep{ecco_consortium_synopsis_2021,ecco_consortium_ecco_2023,tseng_distribution_2025}. A major difference between the 3-D cases and the previous 1-D cases is that spatially dependent growth and decay of algae are considered: the seawater temperature and the light irradiation related to growth are now taken from the ECCOv4r5 ocean state estimate; the mortality and metabolism rates follow \citet{levy_large-scale_2012}, and the grazing rate is now spatially dependent on the background zooplankton concentration from the Darwin Project output \citep{follows_emergent_2007,follows_modeling_2011,dutkiewicz_modeling_2009}.

In earlier survey studies, the ``mortality-grazing rate'' contains high uncertainty, with reported values ranging from $5-88 \%$ day$^{-1}$ \citep{levy_large-scale_2012,calbet_phytoplankton_2004,obayashi_growth_2002}, because the grazing rate depends on the background zooplankton abundance \citep{richon_zooplankton_2022}. We choose the minimal mortality-grazing rate available as the base mortality rate \citep{levy_large-scale_2012}, which better reflects the low-grazing condition in the subtropical oceans \citep{richon_zooplankton_2022}. The remaining explicit grazing is calculated using the predator-prey model and is spatially dependent on the background zooplankton abundance. The resulting mortality and grazing rates sum up to be in the range of $5-85 \%$ day$^{-1}$ for the global ocean, which is consistent with the range of earlier reported values. The specific values of the parameters and treatment on the grazing term in 3-D cases are documented with more detail in \ref{append:parameters}.

\begin{figure}[h!]
\centering
\includegraphics[width=1.\textwidth]{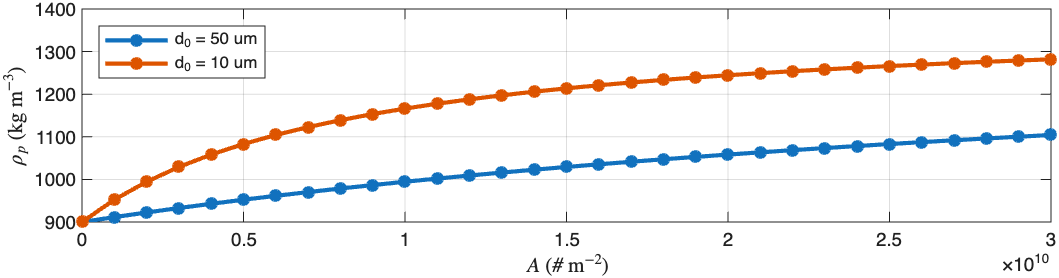}
\caption{The dependency of the effective density $\rho_p$ on the attachment $A$ for different plastic cores. (a) For a core with $\rho_0=900$ kg m$^{-3}$ and $d_0=50\ \mu$m, (b) for a core with $\rho_0=900$ kg m$^{-3}$ and $d_0=10\ \mu$m.}\label{fig:attach_property_cores}
\end{figure}

The dependency of $\rho_p$ on $A$ for the two selected plastic core sizes, $50$ $\mu$m and $10$ $\mu$m, are shown in figure \ref{fig:attach_property_cores}. For both particles, when $A>0$, $\rho_p$ increases monotonically with $A$ because algae cells have a density higher than commonly used plastic materials ($\rho_0<\rho_a$). With the same core density $\rho_0$ and attachment $A$, the effective density of smaller particles is more sensitive to biofouling (figure \ref{fig:attach_property_cores}).

\subsubsection{50 $\mu$m particles}
\label{sec:3D-50}
\begin{figure}[h!]
\centering
\includegraphics[width=1.\textwidth]{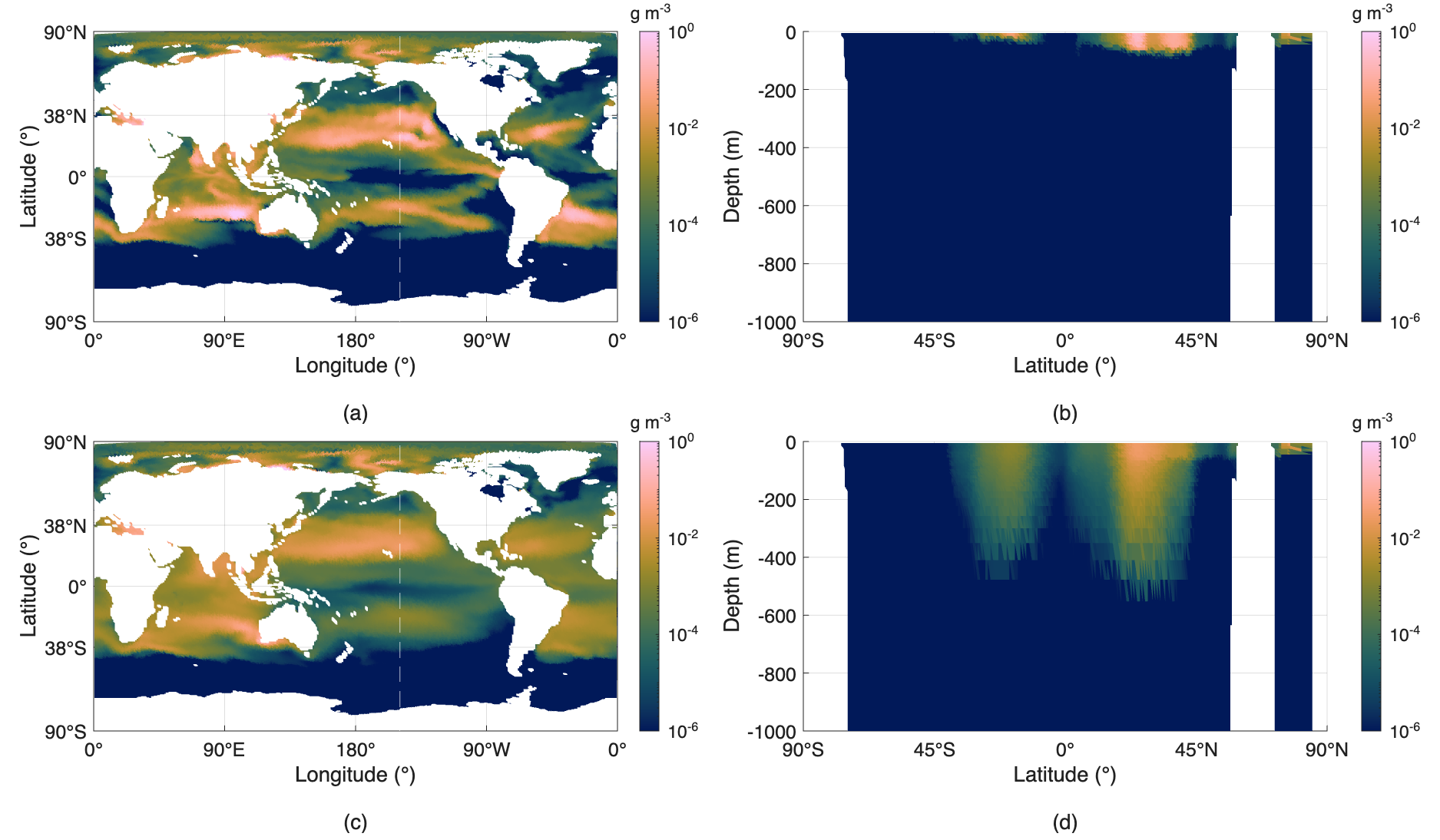}
\caption{The concentration of $50$ $\mu$m MPs (a) on the ocean surface when they are non-biofouled (case 3D-CL-50); (b) on a transect at $150\degree$W when they are non-biofouled (case 3D-CL-50); (c) on the ocean surface when they are biofouled (case 3D-BF-50); and (d) on a transect at $150\degree$W when they are biofouled (case 3D-BF-50).}\label{fig:3D-50-dist}
\end{figure}

We start the discussion with cases 3D-CL-50 and 3D-BF-50 for particles with $50$ $\mu$m core. The results from simulation 3D-CL-50 (with biofouling turned OFF) are shown in figures \ref{fig:3D-50-dist}a and \ref{fig:3D-50-dist}b. At the sea surface (figure \ref{fig:3D-50-dist}a), MP concentration forms clear garbage patches in the five subtropical oceans (Indian, South Pacific, North Pacific, South Atlantic, and North Atlantic) and in the Arctic ocean \citep{mountford_eulerian_2019,tseng_distribution_2025}. In the vertical direction (figure \ref{fig:3D-50-dist}b), most particles are confined in the mixed layer \citep[$\lesssim$ 100 m as shown in the figure;][]{richon_zooplankton_2022,tseng_distribution_2025}. The distribution pattern matches the predictions made by earlier studies for positively buoyant particles \citep{mountford_eulerian_2019,richon_zooplankton_2022}.

For the case 3D-BF-50, the distribution of biofouled particles with $50$ $\mu$m core is shown in figures \ref{fig:3D-50-dist}c and \ref{fig:3D-50-dist}d. At the sea surface (figure \ref{fig:3D-50-dist}c), the five subtropical garbage patches remain visible but become more dispersed (with lower horizontal gradient), and the peak surface concentration is roughly reduced by a factor of $3$ compared to 3D-CL-50, which is consistent with the 1D result (figure \ref{fig:1D-levy}). In terms of the vertical distribution (figure \ref{fig:3D-50-dist}d), the biofouled MPs can sink below the mixed layer. Within the subtropical gyres, biofouled particles can be identified down to $400$ m depth, which is also consistent with the 1D result (figure \ref{fig:1D-levy}). In the Arctic ocean, the garbage patch stays unchanged because the temperature there is not ideal for algae to grow, and the particles can be considered almost clean. In figure \ref{fig:3D-50-dist}d, the vertical distribution of MPs in the Arctic region remains similar to the non-biofouled case.

\begin{figure}[h!]
\centering
\includegraphics[width=1.\textwidth]{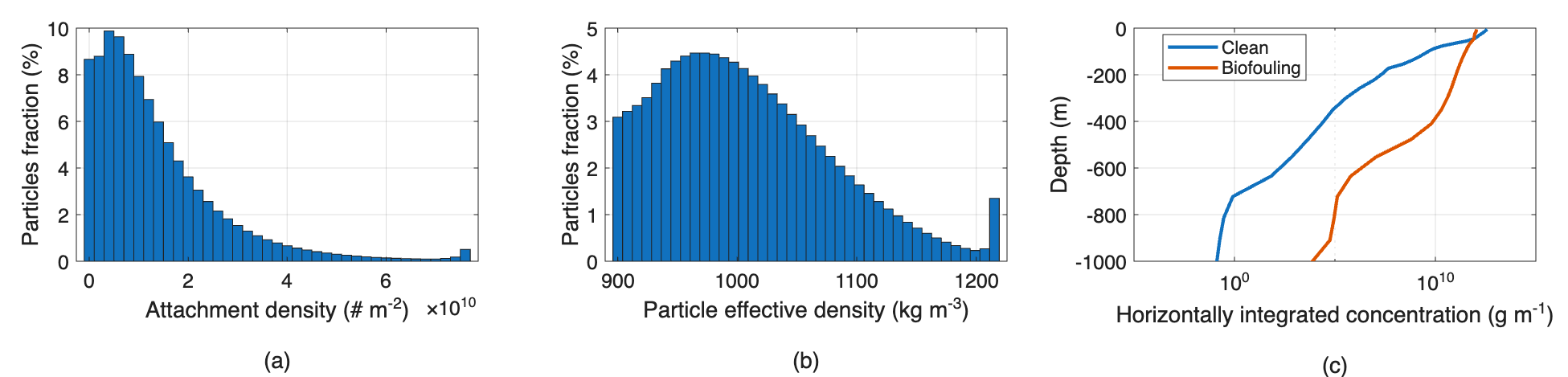}
\caption{(a) The distribution of 3D-BF-50 MPs along the dimension of attachment. (b) The distribution of 3D-BF-50 MPs along the dimension of particle effective density. (c) The horizontally integrated concentration over each level of ocean as a vertical profile.}\label{fig:3D-50-Az}
\end{figure}

To further quantify the effect of biofouling, we show the distribution of particles with respect to the attachment $A$ (figure \ref{fig:3D-50-Az}a). Through \eqref{eq:attch_properties_2}, the distribution is then translated onto its dependence on $\rho_p$ (figure \ref{fig:3D-50-Az}b). This shows the ability of biofouling to modify the property of oceanic MPs, with $\sim 35 \%$ of MPs in the field having a density higher than $1023$ kg m$^{-3}$ (the typical seawater density at the MLD) by the end of the simulation. In terms of the vertical stratification of MPs (figure \ref{fig:3D-50-Az}c), the particle concentration at $400$ m depth can be increased by a factor of $10^5$, indicating a large downward shift of MPs induced by biofouling.

\subsubsection{10 $\mu$m particles}
\label{sec:3D-10}
\begin{figure}[h!]
\centering
\includegraphics[width=1.\textwidth]{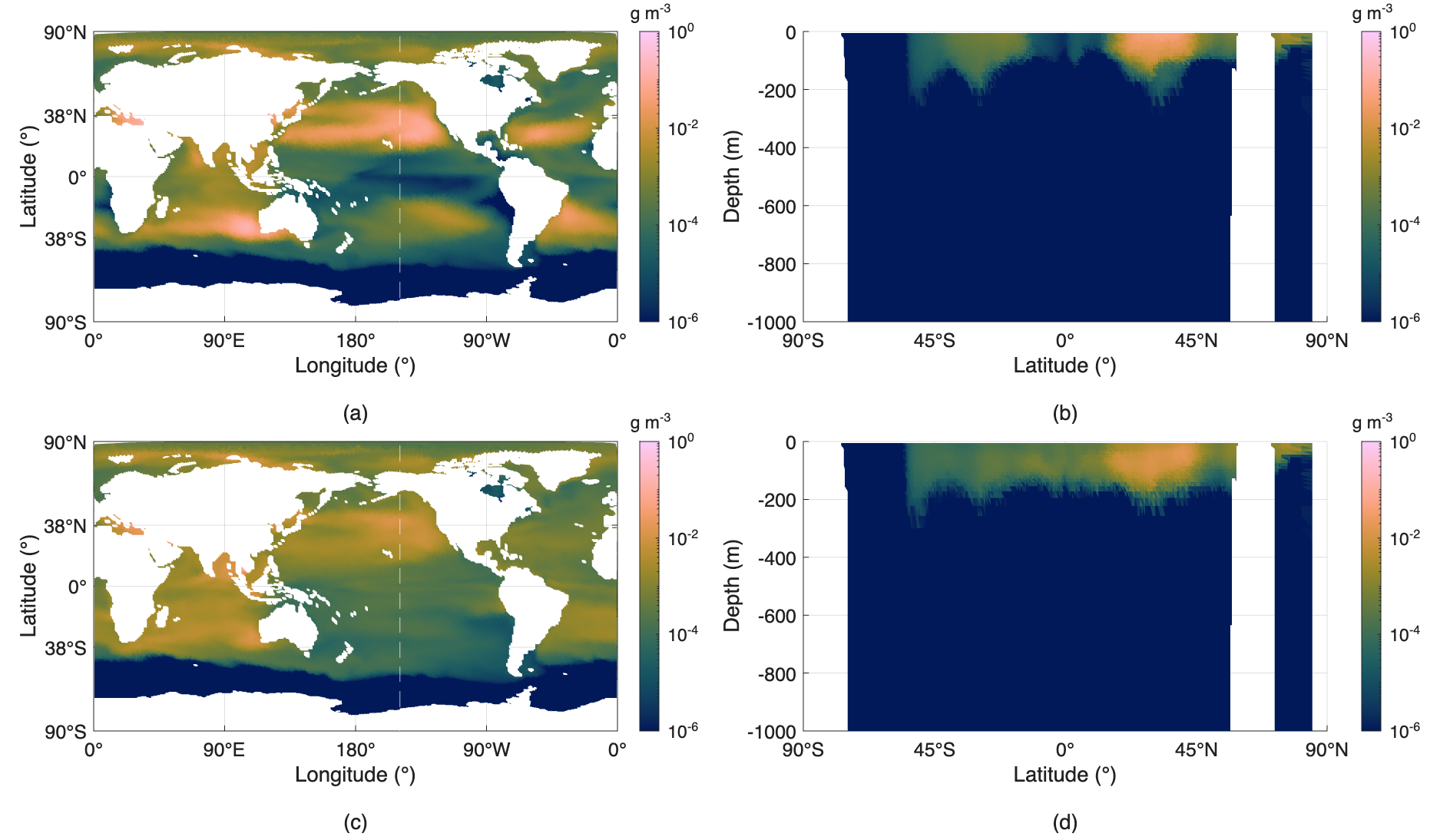}
\caption{The concentration of $10$ $\mu$m MPs (a) on the ocean surface when they are clean (case 3D-CL-10); (b) on a transect at $150\degree$W when they are clean (case 3D-CL-10); (c) on the ocean surface when they are biofouled (case 3D-BF-10); and (d) on a transect at $150\degree$W when they are biofouled (case 3D-BF-10).}\label{fig:3D-10-dist}
\end{figure}

We present the results of cases 3D-CL-10 and 3D-BF-10 for particles with $10$ $\mu$m core in figures \ref{fig:3D-10-dist}a and \ref{fig:3D-10-dist}b to show the distribution of clean MPs (case 3D-CL-10) at the surface layer and at a $150\degree$W transect, respectively. The subtropical garbage patches are again reproduced in this case, and most of the particles stay within the mixed layer. More details and discussion of this clean particle case can be found in the original study \citep{tseng_distribution_2025}.
\begin{figure}[h!]
\centering
\includegraphics[width=1.\textwidth]{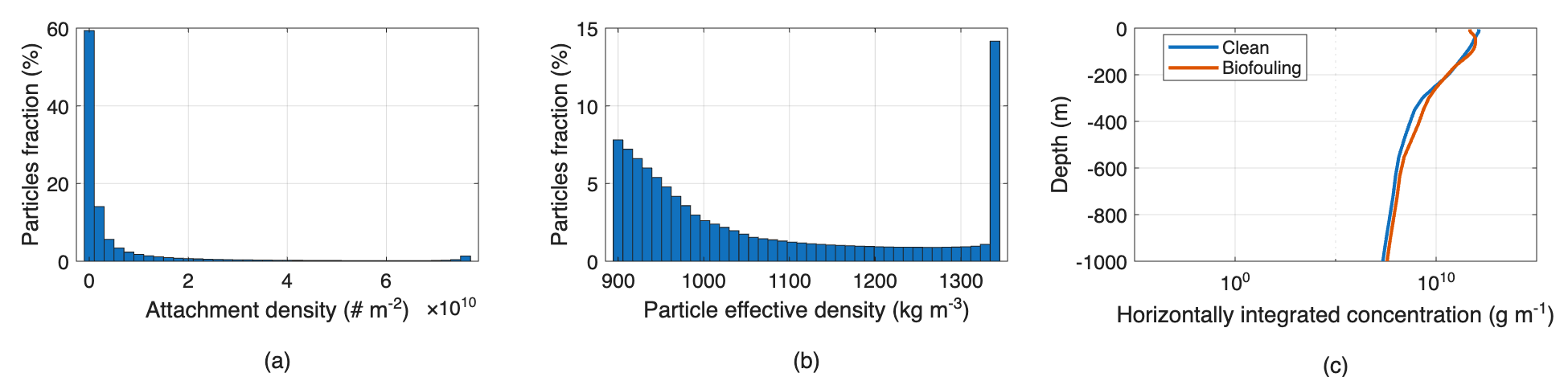}
\caption{(a) The distribution of 3D-BF-10 MPs along the dimension of attachment, with the vertical line dividing buoyant particles to the left and sinking particles to the right. (b) The distribution of 3D-BF-10 MPs along the dimension of particle effective density, with the vertical line dividing buoyant particles to the left and sinking particles to the right. (c) The horizontally integrated concentration over each level of ocean as a vertical profile.}\label{fig:3D-10-Az}
\end{figure}

For the biofouled case 3D-BF-10, the distribution of MPs are plotted in figures \ref{fig:3D-10-dist}c and \ref{fig:3D-10-dist}d. At the sea surface (figure \ref{fig:3D-10-dist}c), the garbage patches are not as clear as they are in the case 3D-CL-10. More specifically, only patches in the Arctic, North Pacific, Indian, and South Atlantic Oceans are still visible, whereas patches in the South Pacific and North Atlantic Oceans can hardly be identified. Also, the downward shift of MPs (figure \ref{fig:3D-10-dist}d) caused by biofouling for $10$ $\mu$m particles ($\sim 100$ m) is weaker than that for $50$ $\mu$m particles ($\sim 400$ m). This can be explained by the stronger sensitivity of effective density on attachment, meaning that $10$ $\mu$m particles would quickly regain buoyancy before sinking to the deeper ocean interior.

The distribution of particles on the dimension of attachment $A$ and the global vertical profile are shown in figure \ref{fig:3D-10-Az}. In this case, $\lesssim 25 \%$ of MPs in the field have a density higher than $1023$ kg m$^{-3}$ (the typical seawater density at the MLD). Compared to the $50$ $\mu$m case, a higher portion of particles remain positively buoyant, because the effective density of $10$ $\mu$m particles is more sensitive to the change of attachment, i.e., once the particle sinks, it is easier to return to the positively buoyant state. Furthermore, as shown in figure \ref{fig:3D-10-Az}c, the concentration of MPs no longer peak at the surface, but at the bottom of the mixed layer. This result aligns with the predictions made in previous Lagrangian model by \citet{lobelle_global_2021}, which dictates that the denser water in the pycnocline provides stronger buoyancy for the particles, thus the vertical profile of biofouled particles is shaped by the MLD.

\subsubsection{1 $\mu$m particles}
\label{sec:3D-1}
For even smaller particles with $1$ $\mu$m core, we have verified that their transport and distribution remain unchanged by biofouling. This is because MPs of this size retain their neutral buoyancy due to negligible terminal velocity \eqref{eq:terminal_velocity} even after being biofouled. Readers are referred to earlier studies which have already shown the distribution of such particles \citep{mountford_eulerian_2019,richon_zooplankton_2022,tseng_distribution_2025}.

\subsection{Physical explanation on the dispersion of garbage patches}
\label{sec:physical_explanation}
We have established that biofouling has a smoothing effect on the surface distribution of MPs. Clearly this is related to the downward shift of MPs by biofouling, which triggers the sinking of particles and reduction of surface concentration as has been discussed. However, that mechanism alone does not explain the dispersion of patches away from the gyre centers. Instead, we show in this section that this dispersion of subtropical garbage patches is further associated with the sub-mixed layer transport. Because biofouled particles can sink to deeper layers, some of them follow an internal transport route, which is not necessarily aligned with the surface circulation. Such vertical inhomogeneity in transport can eventually result in the dispersion of gyres as seen in figures \ref{fig:3D-50-dist}c.

\begin{figure}[h!]
\centering
\includegraphics[width=1.\textwidth]{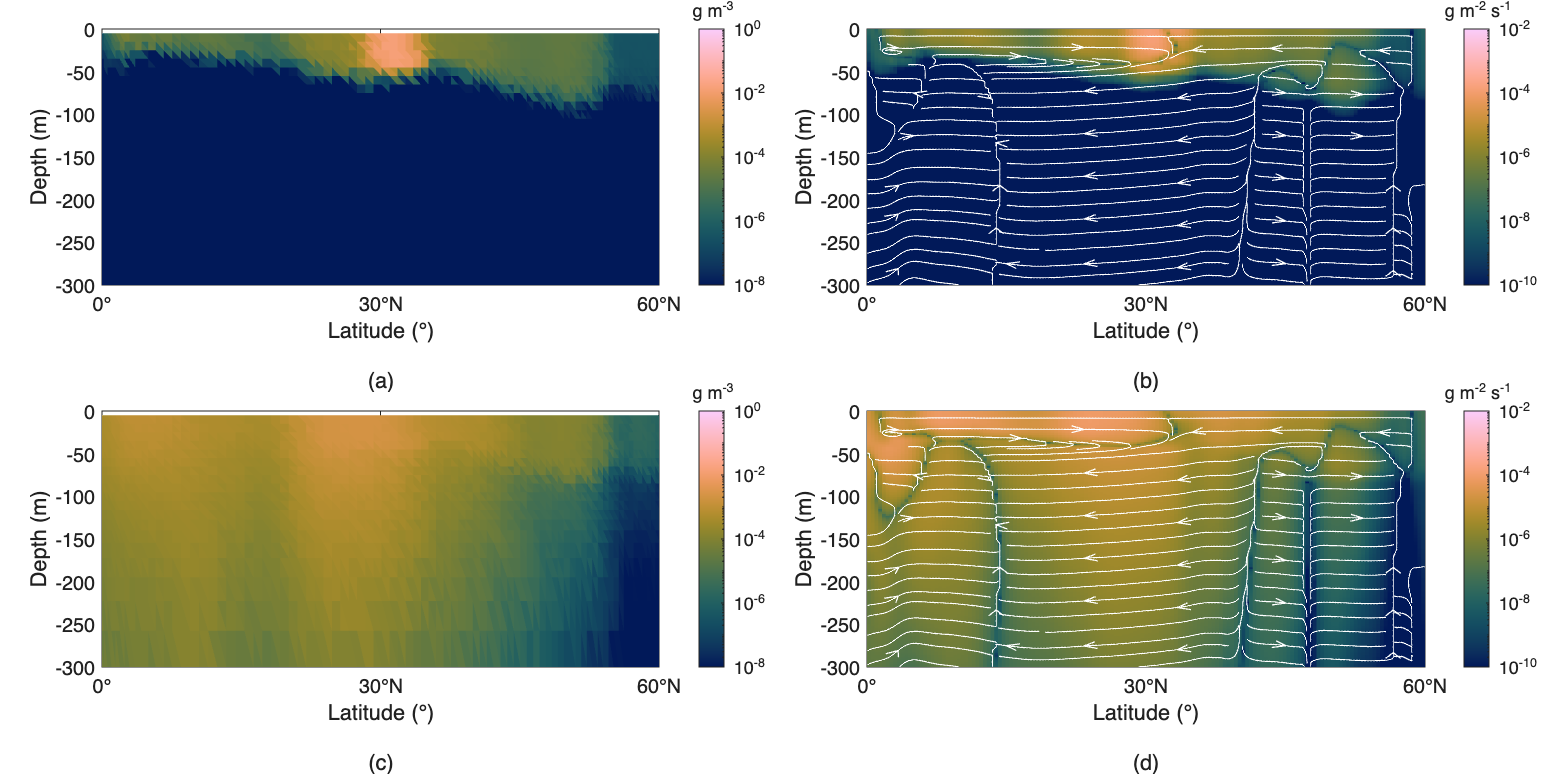}
\caption{The concentration and flux ($\tau v, \tau w$) of MPs in the cases 3AB from the Northern Atlantic, taken on a vertical slice at $30\degree$W. (a) The concentration of $50$ $\mu$m-MPs in case 3D-CL-50 without biofouling. (b) The flux of $50$ $\mu$m-MPs in case 3D-CL-50 without biofouling; the overlaid background represents the magnitude of flux, and the white lines are streamlines of ocean currents. (c) The concentration of $50$ $\mu$m-MPs in case 3D-BF-50 with biofouling. (d) The flux of $50$ $\mu$m-MPs in case 3D-BF-50 with biofouling; the overlaid background represents the magnitude of flux, and the white lines are streamlines of ocean currents.}\label{fig:3AB_flux}
\end{figure}

As an example, we take a close look at the MP distribution in the North Atlantic. In figures \ref{fig:3AB_flux}a--\ref{fig:3AB_flux}d, we show the concentration and flux of MPs on a $30\degree$W transect. As discussed in $\S$\ref{sec:3D-50}, non-biofouled particles in case 3D-CL-50 are confined within the mixed layer (figure \ref{fig:3AB_flux}a). The corresponding flux in figure \ref{fig:3AB_flux}b shows that their meridional transport is dominated by surface currents, which carry clean particles toward the gyre center. In case 3D-BF-50 with biofouling, the concentration field in figure \ref{fig:3AB_flux}c shows that particles are no longer confined to the surface but extend to greater depths. As shown in the flux field in figure \ref{fig:3AB_flux}d, biofouled particles sink below the mixed layer, where sub-mixed layer currents flow in the opposite direction. The biofouled MPs are then carried southward before regaining buoyancy and returning to the surface. This sub-mixed layer transport results in a net flux of MPs directed outward from the gyre center, which disperses the garbage patch and smooths the surface concentration.

\section{Comparison with observations}
\label{sec:compare_obs}

To further assess the performance of our biofouling model, we compare our modeling results with a comprehensive global microplastics dataset \citep{isobe_multilevel_2021}. This dataset is a collection of trawler measurements, which are biased toward larger particles since MPs smaller than about $10$ $\mu$m are unlikely to be captured. Because of the inevitable bias, we only compare our $50$ $\mu$m-simulation (cases 3D-CL-50 and 3D-BF-50) with Isobe's data. To make the model and observation comparable, we follow the three-step workflow: First, zero-concentration data points from Isobe are excluded, because the zero reading might be due to sampling uncertainties \citep{lebreton_numerical_2012}. Second, our modeled concentration field is interpolated onto the sampling locations of Isobe's dataset. Third, each set of data are normalized by their respective geometric means
\begin{equation}
    \tau'_{k} = \dfrac{\tau_k}{\mu_\tau},
\end{equation}
with
\begin{equation}
    \mu_\tau = \left( \prod_{k=1}^{N} \tau_k \right) ^{\frac{1}{N}},
\end{equation}
where $\tau'_k$ is the normalized concentration; $\mu_\tau$ is the geometric mean concentration; $k \in \{1,...,N\}$, with $N$ being the total number of valid sampling points. The geometric mean is used because the data span several orders of magnitude, and the geometric mean is a more robust measure of central tendency than the arithmetic mean.

\begin{figure}[h!]
\centering
\includegraphics[width=0.9\textwidth]{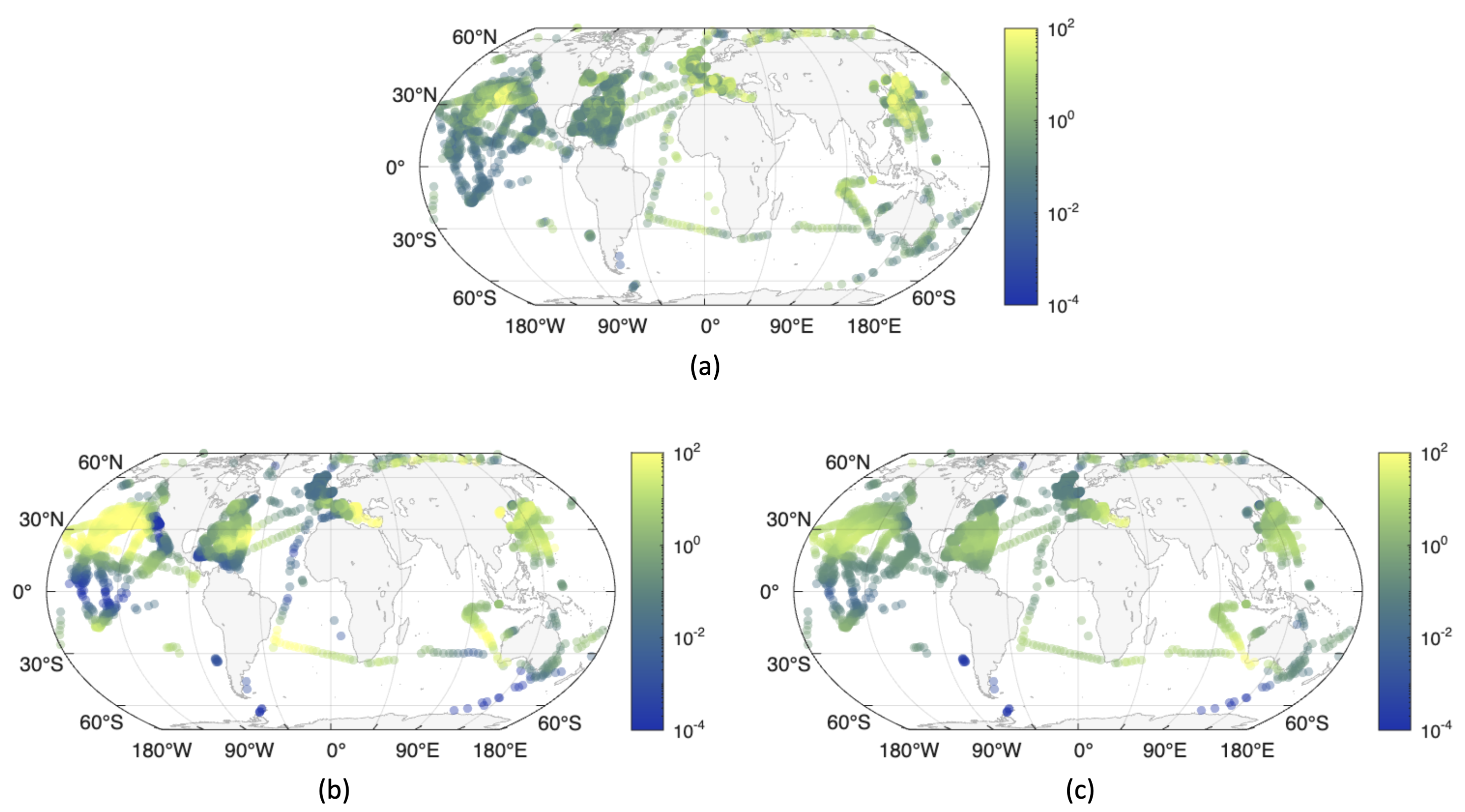}
\caption{Normalized global surface concentration $\tau'$ of large MPs (a) from the Level 1 data of \citet{isobe_multilevel_2021}; (b) the 3D-CL-50 model result; and (c) the 3D-BF-50 model result.}\label{fig:summary_surface}
\end{figure}

We provide in figure \ref{fig:summary_surface} the normalized concentration from Isobe's dataset and from our 3D-CL-50 and 3D-BF-50 modeling results. As shown in figure \ref{fig:summary_surface}a, the samples from \citet{isobe_multilevel_2021} have uneven spatial coverage: nearly half of the samples were collected from the mid-latitude ocean between 30$\degree$N and 60$\degree$N, while the low-latitude Indian Ocean and western Pacific account for only 5$\%$ of all data \citep{isobe_multilevel_2021}. For the case 3D-CL-50 modeling results (figure \ref{fig:summary_surface}b), one distinct feature when compared to the other two figures is that its garbage patches seem more accentuated, i.e. the concentration values are in a wider dynamic range. 

\begin{figure}[h!]
\centering
\includegraphics[width=1.\textwidth]{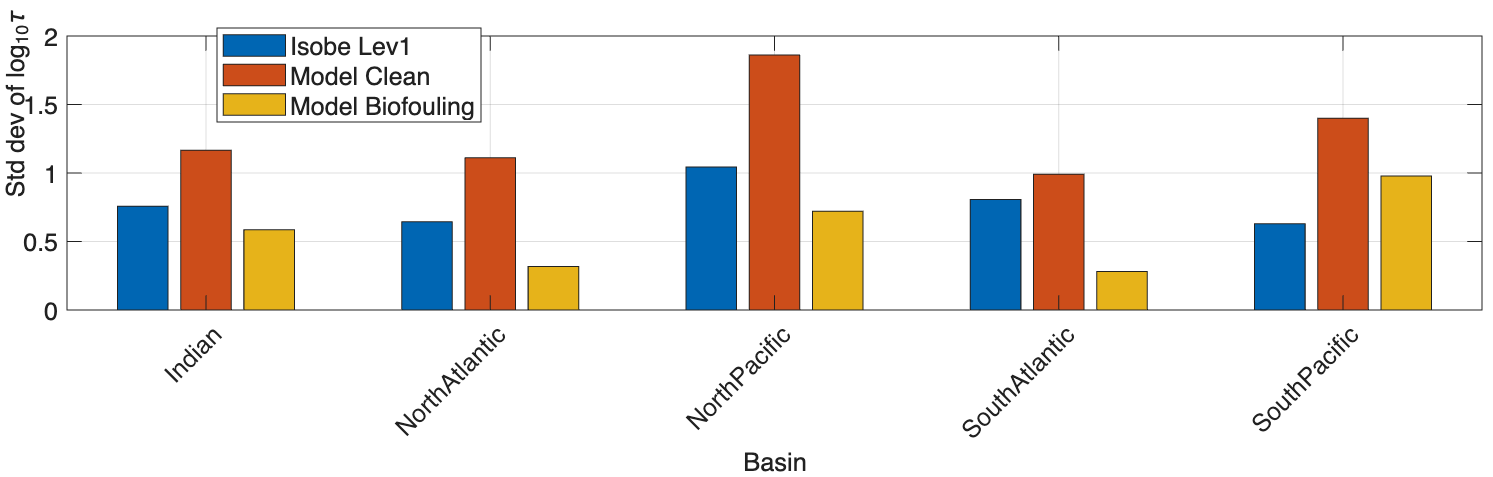}
\caption{Standard deviation of the concentration field from each dataset and divided in the five major ocean basins.}\label{fig:std_each_basin}
\end{figure}

To quantify how biofouling affects the contrast between high- and low-concentration regions, we compute the standard deviation of $\log \tau'$ within each ocean basin, which measures the dynamic range of the surface concentration field in logarithmic space. A higher standard deviation indicates a less uniform concentration field, with more accentuated garbage patches. The results are shown in figure \ref{fig:std_each_basin}. We find that the standard deviation of the modeled MP concentration is significantly reduced by biofouling in all ocean basins, and better aligns with the measured concentration in most ocean basins except the South Atlantic, where over-smoothing is evident. This may reflect insufficient sampling coverage in the South Atlantic to establish statistical significance, as well as uncertainties in the MP sources along the coasts of Africa and South America.

\begin{figure}[h!]
\centering
\includegraphics[width=0.8\textwidth]{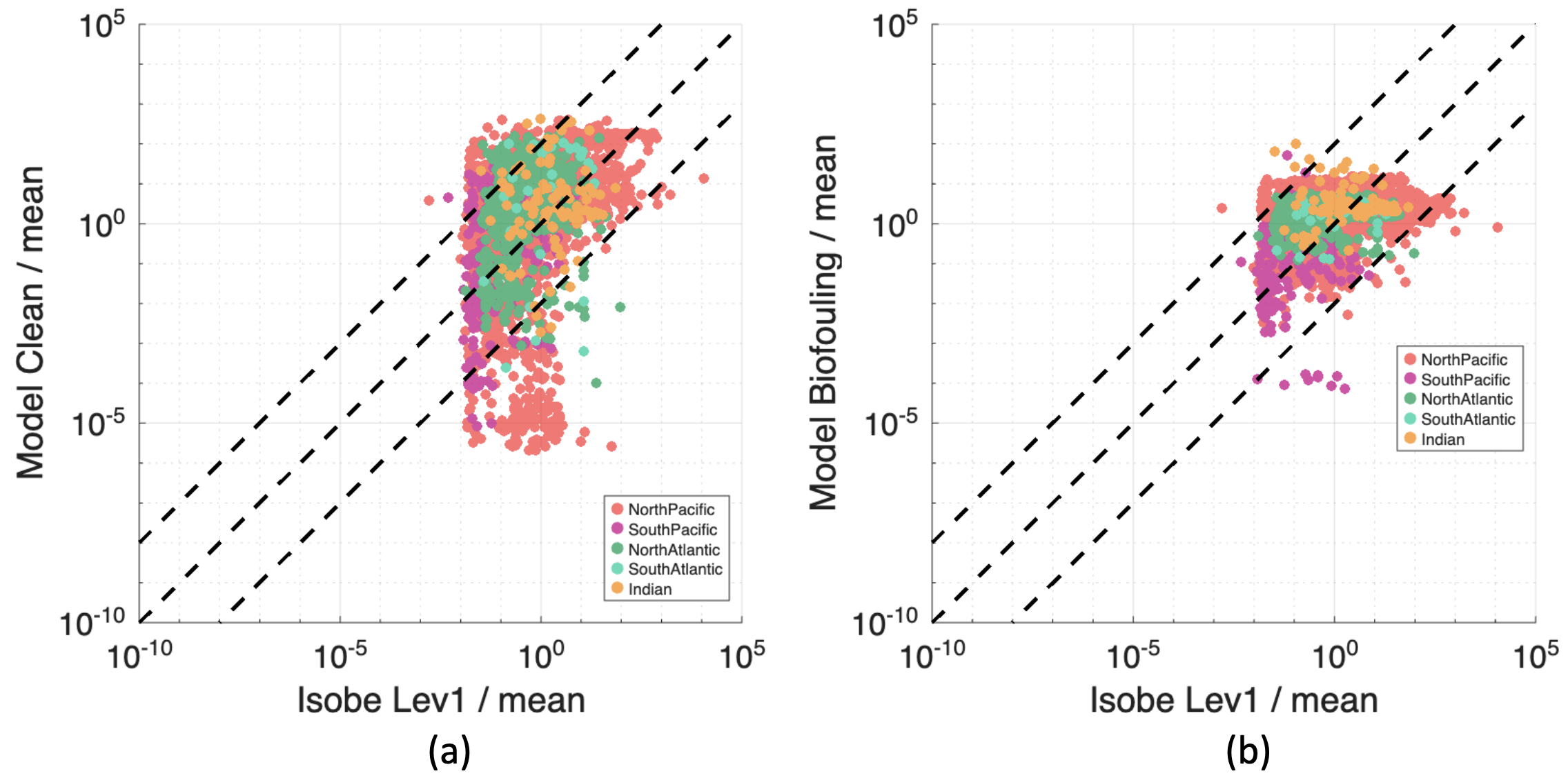}
\caption{Scatter plots of (a) the clean particle concentration versus Isobe's level 1 data; and (b) the biofouled particle concentration versus Isobe's level 1 data. The black dashed lines denote $1:0.01$, $1:1$, and $1:100$.}\label{fig:scatter_normalized}
\end{figure}

In figure \ref{fig:scatter_normalized} we show a direct one-on-one comparison between Isobe's dataset and the modeled concentration. Looking at the dispersion of data points in the scatter plots, it is sensible that the biofouling model concentration is more consistent with observation than the non-biofouled case. As a result, the fraction of outliers (beyond 2 orders of magnitude) is reduced from $25\%$ in 3D-CL-50 to $13\%$ in 3D-BF-50.

\section{Conclusion}
\label{sec:conclusion}

In this study, we quantify the effect of biofouling on the global distribution of MPs, identify the physical mechanisms responsible for the observed changes, and validate our findings against observational data. We present the first 3-D global Eulerian model that fully couples MP transport with biofouling, by augmenting the concentration field with an extra dimension to represent the biomass attachment density. This approach allows us to track how biofouling continuously alters particle properties and the subsequent transport within a global Eulerian framework.

The 1-D simulations demonstrate that biofouling can significantly reshape the vertical distribution of MPs, conditioned on two: (1) the particles must be sufficiently buoyant to remain near the surface where sunlight is available, and (2) the local plankton growth rate must exceed the decay rate, which is typical in the subtropical oceans. The latter condition is determined by the sign of the Floquet exponent (\ref{append:floquet}), which provides a physically grounded criterion for predicting where and when biofouling is dynamically significant.

In the 3-D global simulations, we find that biofouling substantially alters the distribution of large MPs ($\gtrsim 10\ \mu$m). At the sea surface, the subtropical garbage patches become more dispersed with reduced peak concentrations. For $50\ \mu$m particles, the surface concentration is reduced by a factor of $3$, while for $10\ \mu$m particles, the South Pacific and North Atlantic patches disappear entirely. Vertically, biofouled particles are transported below the mixed layer, reaching $500$ m depth. In contrast, small particles ($\lesssim 1$ $\mu$m) remain unaffected by biofouling, as their terminal velocity is negligible and they stay neutrally buoyant regardless of the biomass attached.

We further reveal the mechanism through which biofouling reshapes the surface distribution of MPs. As biofouled particles sink below the mixed layer, they enter the ocean interior where flow direction may not align with surface circulation. These sub-mixed layer currents carry the particles outward from the gyre centers, and when the biofilm eventually decays, the particles regain buoyancy and resurface at locations far from where they originally sank. This sub-mixed layer transport acts as a horizontal redistribution mechanism, dispersing MPs away from the subtropical convergence zones and smoothing the surface concentration field. This finding highlights that the effect of biofouling is not limited to the vertical dimension---it fundamentally alters the pathway by which MPs are transported across the global ocean.

A comparison with the global trawler dataset of \citet{isobe_multilevel_2021} shows that incorporating biofouling partially reduces the model-observation gap. The standard deviation of the concentration is significantly reduced by biofouling across all ocean basins. The fraction of outliers (differing beyond 2 orders of magnitude) is reduced from $25\%$ in the clean-particle simulation to $13\%$ in the biofouled simulation.

Several limitations of the current study highlight key directions for future research. First, the biofouling model considers only a single algal species representative of subtropical oceans, whereas the Darwin Project \citep{follows_modeling_2011} incorporates seven functional groups of algal species across the global ocean with differing cell densities and sizes ($\rho_a, V_a$), as well as distinct growth and decay characteristics. These biogeochemical variations directly modulate parameters such as $\mu_a, Q_{10}, R_{20}, m_a$, and $g_a$ in Kooi’s biofouling equation \eqref{eq:bio-kooi}. By omitting these diverse ecological dynamics, the full complexity and seasonal succession of the plastisphere community remain underrepresented. Second, each simulation is restricted to a single, uniform MP core size and density ($\rho_0, d_0$), neglecting the heterogeneous nature of marine plastic debris. To address this limitation, a follow-up study investigating microplastic fragmentation and the global transport of a broad range of plastic materials is currently underway.

Given the prohibitive computational cost of exploring the full parameter space of plastic materials and biogeochemical parameters, the present study focuses primarily on demonstrating the fundamental role of biofouling in altering the global distribution of microplastics. Ultimately, this work establishes a scalable numerical framework for integrating biofouling dynamics into state-of-the-art global ocean circulation models, such as the Massachusetts Institute of Technology General Circulation Model (MITgcm).

\section*{Data Availability Statement}

To reproduce simulations in this paper, one can follow five procedures briefly summarized below.
\begin{enumerate}
   \item Download MITgcm package.
   \item Download ECCOv4r5 dataset (in particular the forcing and initial conditions for MITgcm to reproduce the ECCOv4r5 dataset).
   \item Download additional code (used in step 4) and inputs (used in step 5) at \url{https://zenodo.org/records/20683542} (recommended) or \url{https://github.com/zizien1019/biofouling_eccov4r5_mitgcm68o} to treat additional terms in \eqref{eq:adv-diff}. Example outputs of simulation 3D-CL-50, 3D-BF-50, 3D-CL-10 and 3D-BF-10 are also provided in the repository.
   \item Compile both the original MITgcm code and additional code in step 3.
   \item Conduct simulations with different inputs on particle properties.
\end{enumerate}

A file uploaded at \url{https://github.com/zizien1019/biofouling_eccov4r5_mitgcm68o/blob/main/Readme.pdf} contains much more detailed step-by-step instructions.

\section*{Acknowledgments}

This research was supported in part by NASA Science Mission Directorate contract 80LARC21DA003 with the University of Michigan.

Portions of this research were conducted with the advanced computing resources provided by Texas A\&M High Performance Research Computing.


\bibliographystyle{elsarticle-harv} 
\bibliography{Zien_paper_2_bibtex_fixed}

\appendix
\input{appendix_draft}

\end{document}

%% file: appendix_draft.tex
\section{Stability analysis}
\label{append:floquet}
We present stability analysis on the property of the biofouling equation \eqref{eq:bio-kooi}. For an initially clean particle, if we let $\mu_a(I,T)-m_a-g_a-Q_{10}^{(T-20)/10} R_{20}=a(t)$, and $\dfrac{\beta_a A_a}{\theta_p}=b$, the equation takes the form
\begin{equation}
x'(t) = a(t) x(t) + b,
\label{eq:bio_simple}
\end{equation}
with an initial condition
\begin{equation}
x(0) = 0.
\label{eq:bio_simple_ic}
\end{equation}

In the ocean, among all the variables related to $a(t)$ (the exponential growth/decay rate), the seawater temperature $T$ varies slowly, and the daylight irradiation $I$ shows periodic daily variation. Thus, $a(t)$ is periodic with a period $T_d=1$ day, $a(t)=a(t+T_d)$, and \eqref{eq:bio_simple} is a linear differential equation with periodic coefficients, that can be analyzed by the Floquet analysis \citep{floquet_sur_1883}.

Let $p(t)=\int_0^{t}a(s)ds$, we can write down the analytical solution to \eqref{eq:bio_simple} at time $t$ and $t+T_d$
\begin{equation}
x(t) = b \cdot  e^{p(t)} \int_0^t e^{-p(s)}ds,
\label{eq:bio_solution}
\end{equation}
\begin{equation}
x(t+T_d) = b \cdot  e^{p(t+T_d)} \int_0^{t+T_d} e^{-p(s)}ds,
\end{equation}

Note that $p(t+T_d)-p(t)=\int_t^{t+T_d}a(s)ds=q$ is a constant because $a$ is $T_d$-periodic. Therefore,
\begin{align}
x(t+T_d)&= b \cdot e^{p(t)} \cdot e^q \cdot \big[ \int_0^t e^{-p(s)}ds + \int_t^{t+T_d} e^{-p(s)}ds \big], \\
        &= e^q \cdot b \cdot e^{p(t)} \int_0^t e^{-p(s)}ds + e^q \cdot b \cdot e^{p(t)} \int_t^{t+T_d} e^{-p(s)}ds, \\
        &= e^q \cdot x(t) + e^q \cdot b \cdot e^{p(t)} \int_t^{t+T_d} e^{-p(s)}ds, \\
        &= e^q \cdot x(t) + e^q \cdot b \cdot e^{p(t)} \cdot e^{-p(t)} \int_0^{T_d} e^{-p(u)}du, \\ 
        &= e^q \cdot x(t) + e^q \cdot b \int_0^{T_d} e^{-p(u)}du. 
\label{eq:bio_derive_1}
\end{align}

Let $e^q = \rho$, and $b \int_0^{T_d} e^{-p(u)}du = C$, which are another two constants, we have
\begin{equation}
    x(t+T_d) = \rho \cdot x(t) + \rho C,
\end{equation}
which is the Poincaré map relating the solution at time $t$ and the solution after one period at time $t+T_d$. By induction, we obtain for all $n \in \mathbb{N}$
\begin{equation}
    x(t+nT_d) = 
    \begin{cases}
        x^* + \rho^n \ [x(t) - x^*] &, \mbox{ if } \quad \rho \neq 1, \\
        x(t)+ nC                    &, \mbox{ if } \quad \rho = 1,
    \end{cases}
\label{eq:bio_poincare}
\end{equation}
where $x^* = \rho C / (1-\rho)$ is the asymptotic value of $x$ in the case when $\rho<1$.

That is, the Lyapunov stability of \eqref{eq:bio-kooi} is consistent with the absolute stability of \eqref{eq:bio_poincare}, which can be determined by the magnitude of $\rho$. Recalling our definition for $\rho$, whether its value is larger than $1$ is determined by the sign of the exponent $q$, which turns out to be the Floquet exponent \citep{floquet_sur_1883} of \eqref{eq:bio-kooi}. The Floquet exponent is defined as the integral of all the coefficients related to exponential growth/decay over one period
\begin{equation}
    q = \int_t^{t+T_d} \big[ \mu_a(I(t'),T) -m_a-g_a-Q_{10}^{(T-20)/10} R_{20} \big] \cdot dt'.
    \label{eq:floquet}
\end{equation}
When $q>0$ it means the solution $A(t)$ is unstable, and the algae can grow indefinitely. When $q<0$ it means the solution $A(t)$ is Lyapunov stable, and the amount of attachment stays finite.

\begin{figure}[h!]
\centering
\includegraphics[width=1.0\textwidth]{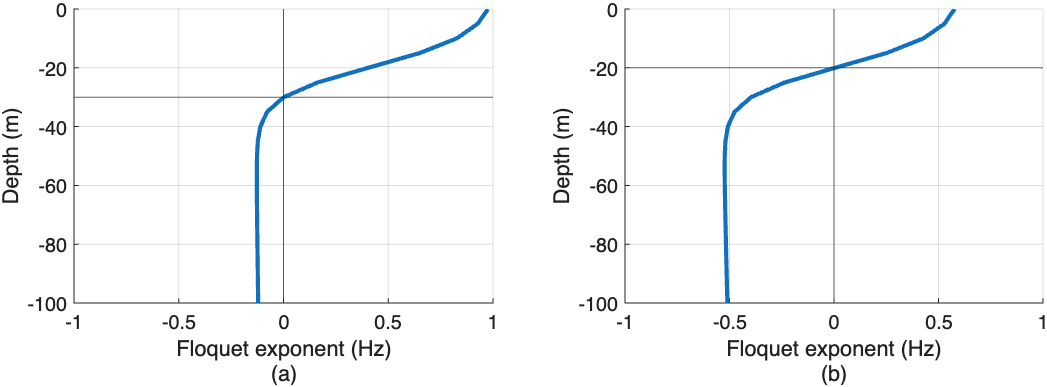}
\caption{The Floquet exponent profile calculated for North Pacific ocean in winter. The biogeochemical parameters are from (a) \citet{levy_large-scale_2012} and (b) \citet{calbet_phytoplankton_2004}. Horizontal lines are added to indicate the zero-passing depth.}\label{fig:floquet}
\end{figure}

Using the empirical seawater temperature and light intensity profiles in North Pacific ocean in winter proposed by \citet{kooi_ups_2017}, we calculate and plot in figure \ref{fig:floquet} the Floquet exponent profiles under two different environmental conditions, differing by the choice of biogeochemical parameters $m_a$, $g_a$, $R_{20}$, and $Q_{10}$ \citep[see Appendix \ref{append:parameters}; ][]{calbet_phytoplankton_2004,levy_large-scale_2012}. In figure \ref{fig:floquet}a, the biogeochemical parameters are taken from \citet{levy_large-scale_2012} and are characterized by lower decay rates ($11\%$ per day in total at $20 \degree$C). In figure \ref{fig:floquet}b, the parameters are taken from \citet{calbet_phytoplankton_2004} and are characterized by higher decay rates ($49\%$ per day in total at $20 \degree$C). 

In figure \ref{fig:floquet}a, the Floquet exponent is positive within $\sim 30$ m of surface ocean. For a positively buoyant particle, the environmental condition would be suitable for the algae to grow, and the particle keeps getting denser and larger until it sinks below $30$ m. For a biofouled particle below $30$ m, it stays negatively buoyant for some time before the attachment decays, until it becomes positively buoyant and floats back to the surface. The traces of such particles collectively result in a change of their vertical distribution.

The predicted Floquet exponent in figure \ref{fig:floquet}b has lower values compared to figure \ref{fig:floquet}a. In this case $q$ is positive within a shallower layer ($\sim 20$ m) of surface, allowing less particles in the unstable region. The weaker positive values and stronger negative values of Floquet exponent suggests that the attachment of floating particles grows slower, and the attachment of interior biofouled particles decays faster. This case represents a harsh environment, where the high decay rate allows a shorter time window for the biofouled particles to sink before they return positively buoyant.

\section{Parameters used in simulations}
\label{append:parameters}

We document all equations and parameters used in the 1-D and 3-D simulations. 
The two subsections below describe the treatment of each term separately for each case.

\subsection{1-D cases (1D-CL, 1D-BF-L\'evy, and 1D-BF-Calbet)}
\label{append:parameters-1D}

For the 1-D cases, the vertical current is weak compared to the terminal velocity, and \eqref{eq:adv-diff} reduces to
\begin{equation}
    \dfrac{\partial \tau}{\partial t} + \partial_z (\tau w_r) = \partial_z(K_{33} \partial_z \tau) + \ \delta (z) \ \delta (A) + Q_{biofouling}.
\label{eq:adv-diff-1D}
\end{equation}

The rising/sinking terminal velocity $w_r$ is computed using Stokes' law
\begin{equation}
    w_r(z,A) = \dfrac{ g \cdot d_p^2 \cdot (\rho_w-\rho_p) }{ 18 \mu },
\end{equation}
where $\rho_w(z)$ is the local seawater density, $\rho_p(A)$ and $d_p(A)$ are the effective density and diameter of the particle, $\mu(z)$ is the dynamic viscosity of seawater, and $g=9.81$ m s$^{-2}$.

The vertical diffusivity $K_{33}(z)$ is computed using the K-profile parametrization (KPP) scheme \citep{large_oceanic_1994}
\begin{equation}
    K_{33}(z) = 
    \begin{cases}
       -h \cdot \kappa \cdot u^* \cdot \phi^{-1} \cdot \sigma \cdot (1-\sigma)^2 &, \mbox{ if } z > h, \\
        0.0002 \text{ m}^2\text{ s}^{-1}               &, \mbox{ if } z \leq h,
    \end{cases}
\label{eq:KPP}
\end{equation}
\begin{table}[h!]
    \centering
    \begin{tabular}{l l c c}
        Symbol & Description & Value & Unit \\
        \hline
        $h$       & Mixed layer depth (MLD) & $-80$ & m \\
        $\kappa$  & von K\'{a}rm\'{a}n constant & $0.4$ & --- \\
        $u^*$     & Friction velocity, $\sqrt{0.2/\rho_w(0)}$ & --- & m s$^{-1}$ \\
        $\sigma$  & Normalized depth, $z/h$ & --- & --- \\
        $\phi$    & Stability function, $1 - 5z/L$ & --- & --- \\
        $L$       & Monin-Obukhov length, $u^{*3}/(\kappa \times 6.3\times10^{-8})$ & --- & m \\
    \end{tabular}
    \caption{Parameters used in the KPP diffusivity scheme \eqref{eq:KPP} for the 1-D simulations.}
    \label{tab:KPP}
\end{table}

The empirical seawater temperature $T$, salinity $S$, density $\rho_w$, and dynamic viscosity $\mu$ profiles used in the 1-D simulations are taken from \citet{kooi_ups_2017}, and are described by the following polynomial fits
\begin{equation}
    T(z) = T_{surf} + (T_{bot} - T_{surf}) \dfrac{|z|^p}{|z|^p + |z_c|^p},
\end{equation}
\begin{equation}
    S(z) = c_1 z^5 + c_2 z^4 + c_3 z^3 + c_4 z^2 + c_5 z + c_6,
\end{equation}
\begin{equation}
\begin{split}
    \rho_w(T,S) = \,& a_1 + a_2 T + a_3 T^2 + a_4 T^3 + a_5 T^4 \\
                   & + b_1\hat{S} + b_2\hat{S}T + b_3\hat{S}T^2 + b_4\hat{S}T^3 + b_5\hat{S}^2 T^2,
\end{split}
\end{equation}
where $\hat{S} = S/1000$. The dynamic viscosity $\mu$ is computed via
\begin{align}
    \mu(T,S) = \,& \left[ 4.2844\times10^{-5} + \frac{1}{0.156(T+64.993)^2 - 91.296} \right] \nonumber \\
    & \times \Big[ 1 + \left(1.541 + 1.998\times10^{-2}T - 9.52\times10^{-5}T^2\right)\hat{S} \nonumber \\
    & \phantom{{}\times\Big[} + \left(7.974 - 7.561\times10^{-2}T + 4.724\times10^{-4}T^2\right)\hat{S}^2 \Big].
\end{align}
\begin{table}[h!]
    \centering
    \begin{tabular}{l l c c}
        Symbol & Description & Value & Unit \\
        \hline
        $T_{surf}$ & Surface temperature & 25 & $\degree$C \\
        $T_{bot}$  & Bottom temperature  & 1.5 & $\degree$C \\
        $p$        & Thermocline shape exponent & 2 & --- \\
        $z_c$      & Thermocline depth & $-300$ & m \\
        $c_1$      & Salinity polynomial coefficient & $9.998\times10^{-17}$ & psu m$^{-5}$ \\
        $c_2$      & & $1.054\times10^{-12}$ & psu m$^{-4}$ \\
        $c_3$      & & $3.997\times10^{-9}$  & psu m$^{-3}$ \\
        $c_4$      & & $6.541\times10^{-6}$  & psu m$^{-2}$ \\
        $c_5$      & & $4.195\times10^{-3}$  & psu m$^{-1}$ \\
        $c_6$      & & $3.517\times10^{1}$   & psu \\
        $a_1$      & Density polynomial coefficient & $9.999\times10^{2}$ & kg m$^{-3}$ \\
        $a_2$      & & $2.034\times10^{-2}$  & kg m$^{-3}$ $\degree$C$^{-1}$ \\
        $a_3$      & & $-6.162\times10^{-3}$ & kg m$^{-3}$ $\degree$C$^{-2}$ \\
        $a_4$      & & $2.261\times10^{-5}$  & kg m$^{-3}$ $\degree$C$^{-3}$ \\
        $a_5$      & & $-4.657\times10^{-8}$ & kg m$^{-3}$ $\degree$C$^{-4}$ \\
        $b_1$      & & $8.020\times10^{2}$   & kg m$^{-3}$ \\
        $b_2$      & & $-2.001$              & kg m$^{-3}$ $\degree$C$^{-1}$ \\
        $b_3$      & & $1.677\times10^{-2}$  & kg m$^{-3}$ $\degree$C$^{-2}$ \\
        $b_4$      & & $-3.060\times10^{-5}$ & kg m$^{-3}$ $\degree$C$^{-3}$ \\
        $b_5$      & & $-1.613\times10^{-5}$ & kg m$^{-3}$ $\degree$C$^{-2}$ \\
    \end{tabular}
    \caption{Parameters for the empirical temperature, salinity, and density profiles used in the 1-D simulations \citep{kooi_ups_2017}.}
    \label{tab:TSrho}
\end{table}

The light intensity $I$ decays exponentially with depth due to attenuation by both water and algae
\begin{equation}
    I(z,t) = \max \big[0, I_0 \cdot \cos(2 \pi t) \cdot \exp\!\left( -\int_z^0 \left[ 0.2 + 0.02 \, C_{chl}(z') \right] dz' \right) \big],
\label{eq:light}
\end{equation}
where $I_0 = 1.2\times10^{8}$ $\mu$E m$^{-2}$ day$^{-1}$ is the surface irradiance, and $C_{chl}$ (mg m$^{-3}$) is the chlorophyll-$a$ concentration profile, taken from \citet{kooi_ups_2017} as
\begin{equation}
    C_{chl}(z) = C_{chl,0} \left[ C_b - s|z| + C_{max} \exp\!\left(-\left(\frac{z-z_{max}}{\Delta z}\right)^{\!2}\right) \right],
\label{eq:chla}
\end{equation}
\begin{table}[h!]
    \centering
    \begin{tabular}{l l c c}
        Symbol & Description & Value & Unit \\
        \hline
        $t$          & Model time & 0--50 & day \\
        $I_0$        & Surface irradiance & $1.2\times10^{8}$ & $\mu$E m$^{-2}$ day$^{-1}$ \\
        $C_{chl,0}$  & Baseline chlorophyll scaling & $0.151$ & mg m$^{-3}$ \\
        $C_b$        & Background chlorophyll coefficient & $0.533$ & --- \\
        $s$          & Slope of chlorophyll decay & $1.72\times10^{-3}$ & m$^{-1}$ \\
        $C_{max}$    & DCM peak chlorophyll & $1.194$ & --- \\
        $z_{max}$    & Depth of DCM (upward-positive) & $-92.01$ & m \\
        $\Delta z$   & Width of DCM & $43.46$ & m \\
    \end{tabular}
    \caption{Parameters for the light attenuation \eqref{eq:light} and chlorophyll-$a$ profile \eqref{eq:chla} used in the 1-D simulations \citep{kooi_ups_2017}.}
    \label{tab:light}
\end{table}

The biofouling equation \eqref{eq:bio-kooi} is
\begin{equation}
    A' \equiv \dfrac{dA}{dt} = \dfrac{\beta_a A_a}{\theta_p} + \mu_a(T,I) A - Q_{10}^{(T-20)/10} R_{20} A - m_a A - g_a A.
\tag{\ref{eq:bio-kooi}}
\end{equation}

The encounter kernel rate $\beta_a$ accounts for Brownian motion, differential settling, and shear
\begin{equation}
    \beta_a = 4\pi (D_p + D_a)(r_p + r_a) + \dfrac{1}{2}\pi r_p^2 |w_{r,p} - w_{r,a}| + 1.3 \, \varepsilon^{1/2} \nu^{-1/2} (r_p + r_a)^3,
\label{eq:beta}
\end{equation}
where $D_p$ and $D_a$ are the Brownian diffusivities of the plastic particle and the algae cell respectively, $r_p$ and $r_a$ are their radii, and $w_{r,p}$, $w_{r,a}$ are their terminal velocities.
\begin{table}[h!]
    \centering
    \begin{tabular}{l l c c}
        Symbol & Description & Value & Unit \\
        \hline
        $r_a$        & Algae cell radius & $3.6\times10^{-6}$ & m \\
        $k_B$        & Boltzmann constant & $1.381\times10^{-23}$ & J K$^{-1}$ \\
        $\varepsilon$ & Turbulent dissipation rate & $1.7\times10^{5}$ & m$^2$ s$^{-3}$ \\[4pt]
        $D_p$        & Diffusivity of particle, $k_B(T+273.16)/(6\pi\mu r_p)$ & --- & m$^2$ s$^{-1}$ \\
        $D_a$        & Diffusivity of algae, $k_B(T+273.16)/(6\pi\mu r_a)$ & --- & m$^2$ s$^{-1}$ \\
    \end{tabular}
    \caption{Parameters used in the encounter kernel $\beta_a$ \eqref{eq:beta}.}
    \label{tab:beta}
\end{table}

The ambient algae concentration $A_a$ (cells m$^{-3}$) is computed from the local chlorophyll-$a$ concentration
\begin{equation}
    A_a = \dfrac{C_{chl}}{R_{chl:C}} \cdot \dfrac{1}{m_{cell}},
    \label{eq:aa_convert}
\end{equation}
where $R_{chl:C}$ is the chlorophyll-to-carbon ratio, estimated as $R_{chl:C} = 0.003 + 0.0154 \exp(0.05 T - 0.059 I/10^6)$, and $m_{cell} = 2726 \times 10^{-9}$ mg cell$^{-1}$ is the carbon mass per algal cell.

\begin{table}[h!]
    \centering
    \begin{tabular}{l l c c}
        Symbol & Description & Value & Unit \\
        \hline
        $\mu_{max}$ & Maximum growth rate  & $1.85$ & day$^{-1}$ \\
        $a$         & Initial slope of $P$-$I$ curve & $0.12$ & day$^{-1}$ ($\mu$E m$^{-2}$ day$^{-1}$)$^{-1}$ \\
        $T_{min}$   & Minimum temperature for growth & $-0.2$ & $\degree$C \\
        $T_{opt}$   & Optimal temperature for growth & $26.7$ & $\degree$C \\
        $T_{max}$   & Maximum temperature for growth & $33.3$ & $\degree$C \\
    \end{tabular}
    \caption{Parameters for the algal growth rate $\mu_a(T,I)$ in \eqref{eq:bio_growth_IandT}--\eqref{eq:bio_growth_T}, common to both 1-D and 3-D simulations \citep{bernard_validation_2012,kooi_ups_2017}.}
    \label{tab:growth}
\end{table}

The algal growth rate $\mu_a$ is modeled as a product of a light-dependent optimal growth rate $\mu_{opt}(I)$ and a temperature scaling function $\Phi(T)$
\begin{equation}
\mu_a(T,I) = \mu_{opt}(I) \  \Phi(T),
\label{eq:bio_growth_IandT}
\end{equation}
\begin{equation}
\mu_{opt}(I) = \mu_{max} \dfrac{I}{I + \dfrac{\mu_{max}}{a} \left(\dfrac{I}{I_{opt}}-1\right)^2},
\label{eq:bio_growth_I}
\end{equation}
\begin{equation}
\Phi(T) = \dfrac{(T-T_{max})(T-T_{min})^2}{(T_{opt}-T_{min})\left[(T_{opt}-T_{min})(T-T_{opt})-(T_{opt}-T_{max})(T_{opt}+T_{min}-2T)\right]},
\label{eq:bio_growth_T}
\end{equation}

\subsubsection{Biogeochemical parameters in cases 1D-BF-L\'evy and 1D-BF-Calbet}

The cases 1D-BF-L\'evy and 1D-BF-Calbet differ by the choice of biogeochemical parameters $m_a$, $g_a$, $Q_{10}$, and $R_{20}$. Case 1D-BF-L\'evy follow \citet{levy_large-scale_2012}, while case 1D-BF-Calbet follow \citet{calbet_phytoplankton_2004}. In both studies, the grazing and mortality rates are reported together as the ``mortality-grazing rate'', i.e., the grazing is implicit, because zooplankton data are not available in the 1-D setting
\begin{table}[h!]
    \centering
    \begin{tabular}{l l c c c}
        Symbol    & Description & \citet{levy_large-scale_2012} & \citet{calbet_phytoplankton_2004} & Unit \\
        \hline
        $m_a+g_a$ & Mortality-grazing rate    & $5\%$ & $39\%$ & day$^{-1}$ \\
        $Q_{10}$  & Metabolic $Q_{10}$ factor & $1.88$              & $2.00$               & --- \\
        $R_{20}$  & Metabolism rate at 20$\degree$C & $6\%$ & $10\%$ & day$^{-1}$ \\
    \end{tabular}
    \caption{Biogeochemical parameters for the 1-D simulations. Case 1D-BF-L\'evy use the parameters from \citet{levy_large-scale_2012} and case 1D-BF-Calbet use the parameters from \citet{calbet_phytoplankton_2004}.}
    \label{tab:biogeo_1D}
\end{table}

\subsection{3-D simulations (cases 3D-[CL/BF]-[50/10/1])}
\label{append:parameters-3D}

In the 3-D simulations, the full transport equation \eqref{eq:adv-diff} is solved on the ECCOv4 llc-90 grid. The ocean velocity $\mathbf{u}$, diffusivity $K$, temperature $T$, salinity $S$, seawater density $\rho_w$, and dynamic viscosity $\mu$ are all taken directly from the ECCOv4r5 ocean state estimate \citep{ecco_consortium_synopsis_2021,ecco_consortium_ecco_2023}.

For the biofouling equation \eqref{eq:bio-kooi}, the ambient algae concentration $A_a$ is computed based on \eqref{eq:aa_convert} with $C_{chl}$ taken from the output of the Darwin Project \citep{follows_emergent_2007,dutkiewicz_modeling_2009,follows_modeling_2011}. 

The light intensity at depth is computed using the same attenuation law as in the 1-D case \eqref{eq:light}
\begin{equation}
    I(x,y,z,t) = max \big[0,I_0(x,y,t) \cdot \exp\!\left( -\int_z^0 \left[ 0.2 + 0.02 \, C_{chl}(z,y,z') \right] dz' \right) \big],
    \label{eq:light-3D}
\end{equation}
but here the surface irradiance $I_0$ is no longer fixed---it is taken from the ECCOv4r5 surface shortwave radiation field, which varies in space and time.

The biogeochemical parameters for the 3-D cases follow \citet{levy_large-scale_2012} and are consistent with the case 1D-BF-L\'evy. The grazing rate $g_a$ is now explicitly computed based on the Darwin Project zooplankton output using the predator-prey model.

\begin{table}[h!]
    \centering
    \begin{tabular}{l l c c}
        Symbol & Description & Value & Unit \\
        \hline
        $m_a$    & Mortality rate    & $5\%$ & day$^{-1}$ \\
        $Q_{10}$ & Metabolic $Q_{10}$ factor & $1.88$ & --- \\
        $R_{20}$ & Metabolism rate at 20$\degree$C & $6\%$ & day$^{-1}$ \\
        $g_a$    & Grazing rate (predator-prey model) & $0$--$80\%$ & day$^{-1}$ \\
    \end{tabular}
    \caption{Biogeochemical parameters for the 3-D simulations \citep{levy_large-scale_2012}.}
    \label{tab:biogeo_3D}
\end{table}

In the Darwin Project, the reduction of phytoplankton due to the grazing behaviour of zooplankton is formulated through the predator-prey model
\begin{eqnarray}
    \dfrac{dP}{dt} \bigg|_{grazing} & = - g_{max} \cdot Z \cdot \dfrac{wP}{K_g + wP} & ,
    \label{eq:predator-prey-darwin}
\end{eqnarray}
where $P$ (typically $P<6$ mmol C m$^{-3}$) is the ambient phytoplankton abundance; $g_{max}$ is the maximum grazing efficiency of zooplanktons; $Z$ (mmol C m$^{-3}$) is the ambient zooplankton abundance; $w$ ($0 \leq w \leq 1$) is the predator preference of the phytoplankton species (with zero meaning non-edible and one meaning strong preference); $K_g$ (mmol C m$^{-3}$) is the half-saturation constant of grazing. It is noted that $dP/dt_{grazing} < 0$ only when $Z$, $w$, and $P$ are all positive. This means that grazing only happens when both zooplankton (predator) and phytoplankton (prey) are present, and when the phytoplankton is edible for the zooplankton. In the Darwin Project, plankton species are divided into seven functional groups in total---five of phytoplankton and two of zooplankton. While in the current study we use a simplified model, where all zooplankton species are classified as the predator, and all phytoplankton species are classified as the prey.

To incorporate the predator-prey model into the biofouling equation \eqref{eq:bio-kooi}, we modified \eqref{eq:predator-prey-darwin} with the following approximation

\begin{equation}
    \dfrac{wP}{K_g + wP} \sim \dfrac{wP}{K_g} , \quad \mbox{given} \quad  wP < 3<< 10.2=K_g.
\label{eq:predator-prey-apr}
\end{equation}

In other words, within the typical working range of the model parameters, the reduction of phytoplankton is approximately linear in the ambient phytoplankton concentration
\begin{equation}
    \dfrac{dP}{dt}\bigg|_{grazing} \sim - g_{max} \cdot Z \cdot \dfrac{wP}{K_g}.
\label{eq:predator-prey-mod}
\end{equation}
Assuming the algae cells attached on the microplastic particles are consumed by the zooplankton due to the same grazing mechanism, we model the reduction rate of algae attachment using the similar linear form
\begin{eqnarray}
    g_a \equiv -\dfrac{dA/dt\big|_{grazing}}{A} \equiv -\dfrac{dP/dt\big|_{grazing}}{P} =   \dfrac{g_{max} \cdot Z \cdot w}{K_g}  & ,
    \label{eq:predator-prey-rate}
\end{eqnarray}
which is the expression of the grazing rate we use in \eqref{eq:bio-kooi}.
\begin{table}[h!]
    \centering
    \begin{tabular}{l l c c}
        Symbol & Description & Value & Unit \\
        \hline
        $g_{max}$ & Maximum grazing efficiency & $62.4\%$ & day$^{-1}$ \\
        $w$       & Predator preference & $0.5$ & --- \\
        $K_g$     & Half-saturation constant & $10.2$ & mmol C m$^{-3}$ \\
        $Z$       & Zooplankton abundance & $0$--$13$ & mmol C m$^{-3}$ \\
    \end{tabular}
    \caption{Parameters for the predator-prey grazing model \eqref{eq:predator-prey-darwin}--\eqref{eq:predator-prey-rate} used in the 3-D simulations \citep{follows_emergent_2007,dutkiewicz_modeling_2009,follows_modeling_2011}.}
    \label{tab:predprey}
\end{table}